\documentclass[dvipsnames,12pt]{article}
\usepackage{graphicx}
\usepackage{verbatim}
\usepackage[urlcolor=OliveGreen, linkcolor=OliveGreen, citecolor=OliveGreen, colorlinks=true]{hyperref}
\usepackage{tocbibind}
\usepackage[round]{natbib}
\usepackage[T2A]{fontenc}
\usepackage[utf8]{inputenc}
\usepackage[english]{babel}
\usepackage{enumerate}
\usepackage{euscript}
\usepackage{pdfpages}
\usepackage[all]{xy}
\usepackage{amsmath, amsthm, amsfonts, amssymb, phonetic, t4phonet, minitoc}
\usepackage{thm-restate}
\usepackage{dirtytalk}
\usepackage{soul}

\usepackage[font=small,labelfont=bf]{caption} 
\usepackage[labelformat=simple]{subcaption}    

\usepackage{tikz}
\usetikzlibrary{arrows, patterns, decorations.pathmorphing,backgrounds,positioning,fit,petri}
\usepackage{pgfplots}
\usepackage{pgfplotstable}

\pgfplotsset{every tick label/.append style={scale=0.65}}
\pgfplotsset{compat=1.13}
\pgfplotsset{mystyle/.style={legend style={at={(.95,.97)}}, 
ylabel=Frequency, xlabel=CCEI, 
xmin=0,xmax=1.02, 
ymin=0,ymax=1, 
ybar=2.5pt, 
bar width=2.5pt}}

\pgfplotsset{mystyle2/.style={legend pos=north west, 
ylabel=Frequency, xlabel=CCEI, 
xmin=0,xmax=1.02, 
ymin=0,ymax=1, 
ybar=3.5pt, 
bar width=2.5pt}}

\definecolor{masongreen}{RGB}{30,98,56} 

\usepackage{cleveref}

\usepackage{algorithmicx, float, algcompatible, algorithm, algpseudocode}


\newtheorem{thm}{Theorem}

\newtheorem{defn}{Definition}

\newtheorem{rem}{Remark}

\newcommand{\R}{\mathbb{R}}

\newcommand{\black}{\textcolor{black}}

\usepackage[margin=1in]{geometry} 

\title{A General Revealed Preference Test for Quasilinear Preferences: Theory and Experiments}
\author{Marco Castillo\thanks{Department of Economics, Texas A\&M University; Melbourne Institute, University of Melbourne; IZA (Bonn, Germany), e-mail: \texttt{marco.castillo@tamu.edu}} \and Mikhail Freer\thanks{Department of Economics, University of Essex.  e-mail: \texttt{m.freer@essex.ac.uk} }}

\begin{document}

\maketitle

\begin{abstract}
We provide a generalized revealed preference test for quasilinear preferences. 
The test applies to nonlinear budget sets and non-convex preferences as those found in taxation and nonlinear pricing contexts. We study the prevalence of quasilinear preferences in a laboratory real-effort task experiment with nonlinear wages. The experiment demonstrates the empirical relevance of our test. 
We find support for either convex (non-separable) preferences or quasilinear preferences but weak support for the hypothesis of both quasilinear and convex preferences.\vspace{0.5in}
\end{abstract}

\newpage

\section{Introduction}

It is difficult to overstate the importance of the assumption of quasilinear (QL) preferences in both theoretical and empirical economics. 
The assumption plays a crucial role in mechanism design, the theory of the household, and applied welfare analysis. 
It is, for instance, a necessary assumption for the Revenue Equivalence theorem \citep{myerson1981,krishna2009}, the existence of the truth-revealing dominant strategy mechanism for public goods \citep{green1977}, and the Rotten Kids theorem \citep{becker1974,bergstrom1983}. 
It is also a frequently invoked assumption in applied welfare analysis \citep{domencich1975urban,allcott2015evaluating}. 
It is also a crucial assumption in the empirical literature that uses bunching at kinks and notches of budget sets \citep[see][for a review]{kleven2016bunching}.
The first contribution of our paper is to provide a new test for QL preferences with many empirical applications. The second contribution is to provide the first empirical test of QL preferences in a laboratory setting using nonlinear budget sets. The third contribution is to show the empirical relevance of QL, albeit not convex preferences.

\paragraph{Contribution}
We provide criteria for a set of observed choices on nonlinear budget sets to be consistent by quasilinear preferences that are not necessarily convex. 
Our criteria are based on the observation that when preferences are QL, they must also satisfy the property of \textit{cyclical monotonicity} \citep[see, e.g.,][]{rochet1987necessary}. 
Our contribution demonstrates that cyclical monotonicity holds for a larger class of choice environments than previously established. 
As a special case, we show that \cite{forges2009}'s generalization of revealed preferences tests to nonlinear sets extends to tests of QL. This allows testing for QL preferences in some strategic environments. When the choice sets are linear, however, the condition we identify is equivalent to the definition of \textit{cyclical monotonicity} in \cite{brown2007}. These conditions are necessary and sufficient for choices to be rationalized by a QL utility function. We provide examples that show why it is essential to consider nonlinear budget sets.

    We conduct a real-effort labor supply experiment with nonlinear wages.
    We allow subjects to choose the terms of their contracts, where a contract specifies a wage and the number of tasks to complete.
    We generate nonlinear convex budgets sets to test for QL preferences and the convexity of preferences. 
    {We find that 77 percent of subjects are consistent with locally non-satiated utility (LNU), of those 86 percent are also consistent with QL and 82 percent are consistent with non-separable convex preferences and 56 are consistent with QL convex preferences.}\footnote{
    {
    This is the percent are participants for whom we can reject random behavior at a $5\%$ significance level. Details of these calculations are in Section \ref{sec:ExperimentalResults}.
    }
    }
    While both assumptions of {convex} or QL preferences have significant empirical support, less is so for QL convex preferences.

\paragraph{Related Literature}
Revealed preference theory is attractive because of its robustness to functional form assumptions regarding preferences.
Beginning with the work of \cite{richter1966} and \cite{afriat1967}, revealed preference theory has been used to test both individual and collective decision-making \citep[see][for a comprehensive overview of the results]{chambers2016revealed}. 
It has also been used to test theories of consumer behavior in the lab.
Applications include tests for social \citep[e.g.,][]{andreoni2002giving,kariv2007,porter2015love}, risk \citep[e.g.,][]{choi2007consistency}, and time preferences \citep[e.g.,][]{andreoni2012estimating}.

Recent experimental research has used real-effort task to study time-preferences \citep[][]{augenblick2015working} and self-control \cite{toussaert2018eliciting}. Our experiment shows that such an environment is useful to test other properties of labor-leisure decisions.
The labor supply setting is particularly relevant in the analysis of household decisions \citep[see e.g.][]{cherchye2012married,cherchye2017household}. Our experimental design can be adapted to study individual and group labor supply behavior.

Recently, there has been interest in revealed preference tests for QL preferences in the context of linear budgets. 
\cite{brown2007} propose a revealed preference test for the case of concave preferences, while 
\cite{nocke2017quasi} discuss the corresponding integrability problem without assuming concavity. 
\cite{cherchye2015utility} extends \cite{brown2007}'s test to generalized QL while maintaining the concavity of the utility and the linearity of budgets.
\cite{allen2018assessing} provide a measure for QL misspecification and use scanner data to evaluate how this misspecification varies with the level of aggregation of the data.
They find that while quasilinearity fails at the individual level, it represents the data well from the perspective of a representative agent.
\cite{chambers2017characterization} show that QL preferences in the setting of combinatorial demand (with linear pricing) are equivalent to the law of demand, a condition that is simpler than cyclical monotonicity. 
As we will discuss, our approach provides the least restrictive test for QL preferences, and our experiment provides direct evidence that quasilinearity has empirical relevance at the \textit{individual} level. 
Our experiments, therefore, complement these approaches nicely.

\paragraph{Structure}
The remainder of this paper is organized as follows. 
Section 2 presents our theoretical framework,
Section 3 presents our experimental design
and Section 4 presents the empirical results.
Section 5 concludes the paper.

\section{Theoretical Framework}
In this section, we discuss the necessary and sufficient conditions for choices to be rationalized {with} QL {preferences}.

Consider a space of alternatives $Y= X \times \R$, where $X\subseteq \R^n_+$.
We denote an element of this set as $(x,m)\in Y$, where $x\in X$ and $m\in \R$.
Intuitively, $x$ is a bundle of consumption goods, and $m$ is an amount of money {left after purchasing $x$}.
In a labor supply setting, as in our experiment, $x$ is leisure time {(total time budget minus the time needed to complete the task)}, and $m$ is the compensation for the task.
In what follows, we will use the consumption analogy due to its greater generality.
A QL {utility function} takes the following form:
$$
v(x,m) = u(x) + m,
$$
where $u: X \rightarrow \R$ is a real valued sub-utility function.

Let $E=((x^t,m^t),B^t)_{t=1}^T$ be a consumption experiment, where $(x^t,m^t)$ is the bundle chosen from budget $B^t\subseteq Y$ and $T$ is the total number of decisions made.
We assume that budgets are downward closed\footnote{
    That is, if $y\in B^t$, then $y'\in B^t$ for every $y'\leq y$.
}
and can be described by a continuous and decreasing function $\mu^t: X \rightarrow \R$,\footnote{
  A function $\mu: X\rightarrow \R$ is decreasing if $x\ge x'$ implies $\mu(x')\ge \mu(x)$ and $x > x'$ implies $\mu(x') > \mu(x)$.
} 
such that
$$
B^t = \lbrace (x,m)\in Y: \mu^t(x)\ge m  \rbrace.
$$
The assumptions we make guarantee that the border of the budget set is downward sloped and represents a closed set. 
These assumptions allow for nonlinear budget sets.

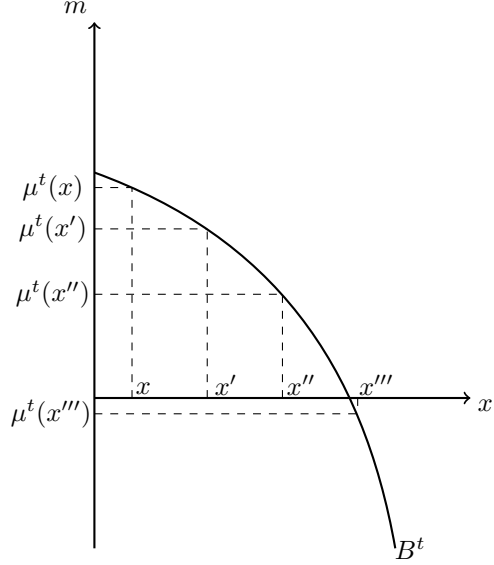
\begin{figure}[htb]
\centering
\begin{tikzpicture}
\draw[thick,->] (0,-2) -- (0,5);
\draw[thick,->] (0,0) -- (5,0);
\node at (5.2,-.1) {\footnotesize$x$};
\node at (-.25,5.2) {\footnotesize$m$};
\draw [thick]   (0,3) to[out=-20,in=100] (4,-2);
\node at (4.2,-2) {\footnotesize$B^t$};

\draw[dashed] (.5,0) -- (.5,2.8);
\node at (.65,.125) {\footnotesize$x$};
\draw[dashed] (0,2.8) -- (.5,2.8);
\node at (-.55,2.8) {\footnotesize$\mu^t(x)$};

\draw[dashed] (1.5,0) -- (1.5,2.25);
\node at (1.75,.175) {\footnotesize$x'$};
\draw[dashed] (0,2.25) -- (1.5,2.25);
\node at (-.55,2.25) {\footnotesize$\mu^t(x')$};

\draw[dashed] (2.5,0) -- (2.5,1.38);
\node at (2.75,.175) {\footnotesize$x''$};
\draw[dashed] (0,1.38) -- (2.5,1.38);
\node at (-.55,1.38) {\footnotesize$\mu^t(x'')$};

\draw[dashed] (3.5,0) -- (3.5,-.21);
\node at (3.75,.175) {\footnotesize$x'''$};
\draw[dashed] (0,-.21) -- (3.5,-.21);
\node at (-.575,-.21) {\footnotesize$\mu^t(x''')$};

\end{tikzpicture}
\caption{Obtaining $\mu^t(x)$.}
\label{fig:GettingM}
\end{figure}

\noindent 
Figure \ref{fig:GettingM} illustrates how one can construct $\mu^t(x)$ using the depicted border of the budget set.
Note that $\mu^t(x)$ can be negative; indeed, for a high enough value of $x$, it \textit{will} be negative.
For the case of linear budgets, $\mu^t(x) =\frac{I-p_x x}{p_m}$, where $I$ stands for income, $p_x$ the price of the good(s) $x$, and $p_m$ the price of money.
We assume that $m^t=\mu^t(x^t)$ in order to avoid a violation of monotonicity in $m$. Since the budget is downward closed for every $x\in X$, $(x,\mu^t(x))\in B^t$, we allow $\mu^t(x)$ to be negative, as this technical assumption simplifies the analysis. 
Intuitively one can think of $\mu^t(x)$ as {the amount of money left after purchasing bundle} $x\in X$ given budget $B^t$.
Note that we require $\mu^t(x)$ to be defined for every $x\in X$.
If we consider $m$ to be money, then a negative amount of money can be interpreted as borrowing, and budgets being downward closed can be interpreted as the absence of a binding liquidity constraint.

\begin{defn}
A consumption experiment $E=((x^t,m^t),B^t)_{t=1}^T$  can be \textbf{rationalized with QL preferences} if there is a function $v(x,m)=u(x)+m$ such that

$$
u(x^t)+m^t \geq u(x)+m \ \text{ for every } (x,m)\in B^t.
$$
\end{defn}

\begin{defn}
A consumption experiment $E=((x^t,m^t),B^t)_{t=1}^T$ satisfies \textbf{cyclical monotonicity} if for every \black{sequence of observations} ${k_1},\ldots, {k_n}\in \{1,\ldots, T\}$,
$$
\sum\limits_{k=1}^{n} \mu^{k_j}(x^{k_j})-\mu^{k_{j+1}}(x^{k_{j}})\geq 0,
$$

\noindent where $k_{n+1}=k_1$.
\end{defn}

\noindent Cyclical monotonicity was introduced by \cite{rockafellar1970convex}, and its importance in characterizing QL was established by \cite{rochet1987necessary}, \cite{brown2007} and \cite{nocke2017quasi}. The version of cyclical monotonicity we use is equivalent to the standard definition up to reordering the terms.

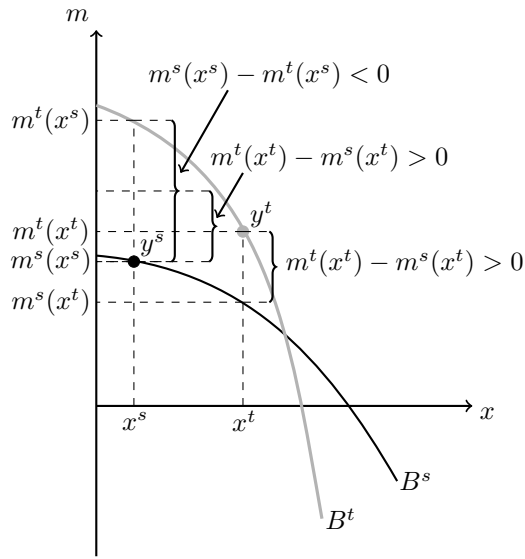
\begin{figure}[htb]
\centering
\begin{tikzpicture}
\draw[thick,->] (0,-2) -- (0,5);
\draw[thick,->] (0,0) -- (5,0);
\node at (5.2,-.1) {\footnotesize$x$};
\node at (-.25,5.2) {\footnotesize$m$};
\draw [thick]   (0,2) to[out=-5,in=120] (4,-1);
\node at (4.25,-1) {\footnotesize$B^s$};
\draw [very thick, gray!60]   (0,4) to[out=-20,in=100] (3,-1.5);
\node at (3.25,-1.5) {\footnotesize$B^t$};

\draw [fill=black] (.5, 1.92) circle (.075cm);
\draw [gray!60, fill=gray!60] (1.95, 2.32) circle (.075cm);
\node at (.75,2.14) {\footnotesize$y^s$};
\node at (2.2,2.55) {\footnotesize$y^t$};

\draw[dashed] (.5,0) -- (.5,3.8);
\draw[dashed] (.5,1.92) -- (0,1.92);
\draw[dashed] (.5,3.8) -- (0,3.8);
\node at (-.6,1.92) {\footnotesize $m^s(x^s)$};
\node at (-.6,3.8) {\footnotesize $m^t(x^s)$};
\node at (.52,-.2) {\footnotesize $x^s$};

\draw[dashed] (1.95,0) -- (1.95,2.32);
\draw[dashed] (1.95,2.32) -- (0,2.32);
\draw[dashed] (1.95,1.38) -- (0,1.38);
\node at (-.6,2.32) {\footnotesize $m^t(x^t)$};
\node at (-.6,1.38) {\footnotesize $m^s(x^t)$};
\node at (1.97,-.2) {\footnotesize $x^t$};

\draw[dashed] (1.95,1.38)--(2.3,1.38);
\draw[dashed] (1.95,2.32) -- (2.3,2.32);
\draw[decorate,decoration={brace,mirror}, thick] (2.3,1.38) -- (2.3,2.32);
\node at (4.1,1.9) {\footnotesize$m^t(x^t)-m^s(x^t)>0$};

\draw[dashed] (.5,1.92) -- (1,1.92);
\draw[dashed] (.5,3.8) -- (1,3.8);
\draw[decorate,decoration={brace,mirror}, thick] (1,1.92) -- (1,3.8);
\draw[thick,<-] (1.12,3.1) -- (1.75,4.2);
\node at (2.3,4.4) {\footnotesize$m^s(x^s)-m^t(x^s)<0$};

\draw[decorate,decoration={brace,mirror}, thick] (1.5,1.92) -- (1.5,1.92+2.32-1.38);
\draw[dashed] (1.5,1.92+2.32-1.38) -- (0,1.92+2.32-1.38);
\draw[dashed] (1.5,1.92) -- (0,1.92);
\draw[thick,<-] (1.6,2.5) -- (1.95,3.1);
\node at (3.1,3.3) {\footnotesize$m^t(x^t)-m^s(x^t)>0$};

\end{tikzpicture}
\caption{Data that are inconsistent with QLU rationalization.}
\label{fig:Violation}
\end{figure}

Figure \ref{fig:Violation} shows an example of a violation of QL preferences \black{although this set of data is rationalizable with a locally non-satiated utility function (LNU)}. To see this, suppose that the agent makes decisions according to QL {preferences}.
{
Since $y^s$ is chosen from $B^s$, it must be the case that  the agent prefers $y^s = (x^s, \mu^s(x^s))$ to $(x^t,\mu^s(x^t))$, then using quasliniearity we can add $[\mu^t(x^t)-\mu^s(x^t)]$ to $y^s$ and to $(x^t, \mu^s(x^t))$ preserving the relation between them.
Thus, $(x^s,\mu^s(x^s)+[\mu^t(x^t)-\mu^s(x^t)])$ is preferred to $y^t = (x^t, \mu^s(x^t)+[\mu^t(x^t)-\mu^s(x^t)])$.
Given that $\mu^t(x^t)-\mu^s(x^t) \leq \mu^s(x^s)-\mu^t(x^s)$ we can conclude that $(x^s,\mu^s(x^s)+[\mu^t(x^t)-\mu^s(x^t)])$ is in the interior of $B^t$.
Thus $y^t$ is strictly revealed preferred to $(x^s,\mu^s(x^s)+[\mu^t(x^t)-\mu^s(x^t)])$, which is a contradiction.
}
We now establish the main result for this section, whose proof is provided in Appendix A.

\begin{thm}
\label{WeakRepresentation}
A consumption experiment $E=((x^t,m^t),B^t)_{t=1}^T$ can be rationalized {with} QL {preferences} if and only if it satisfies cyclical monotonicity.
\end{thm}

\noindent 
We provide the intuition for the proof before we proceed further. 
The main step of the proof is to assign utility numbers to observed choices that are then extended to the whole choice space using techniques developed in \cite{afriat1967}.
Since the necessity of cyclical monotonicity is clear, we concentrate on why cyclical monotonicity allows finding the proper utility values. We start by defining some graph-theoretic notions needed for the argument.

Consider a weighted directed complete graph\footnote{
A graph is said to be complete if there is an edge between every pair of vertexes.
The construction we use is similar to that in \cite{piaw2003afriat}.
} 
where each vertex corresponds to an observation.
The weight assigned to each edge corresponds to the difference in the relative cost of bundles $x^t$ and $x^s$ in the budget $B^t$, that is  $\mu^t(x^t)-\mu^t(x^s)$.
Cyclical monotonicity guarantees that this graph does not contain negative cycles.
Hence, we can assign a utility number to the point $x^t$ that is equal to the weight of the shortest walk passing through $x^t$ at the first transition.
This construction guarantees that the chosen point is better than any other point in the budget. For this, we need to show that chosen values $u(x^t)$ satisfy
$$
u(x^t)+\mu^t(x^t)\geq u(x^s)+\mu^t(x^s) \Leftrightarrow u(x^t) - u(x^s) \geq  \mu^t(x^s)-\mu^t(x^t).
$$
By construction $u(x^s)$ cannot exceed $u(x^t) + (\mu^t(x^s)-\mu^t(x^t))$, since both $u(x^s)$ and $u(x^t)$ are defined as shortest walks.
This observation guarantees that QL rationalizability is satisfied for the space of chosen points. Appendix A shows that $u(x)$ can be assigned using the technique developed in \cite{afriat1967}.

Before concluding this section, we make several remarks that illustrate the difference between our test for quasilinearity and existing ones.
We start from showing that the assumption of convexity of preferences is restrictive, and show how to modify the test by \cite{brown2007} for the setting we consider.

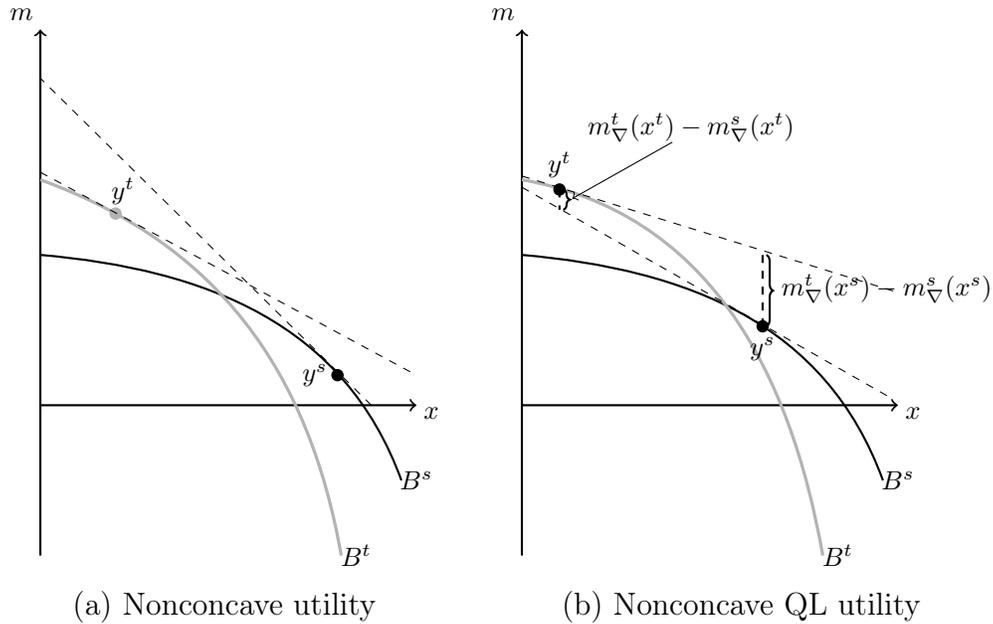
\begin{figure}[htb]
\centering
\begin{tabular}{cc}
\begin{tikzpicture}

\draw[thick,->] (0,-2) -- (0, 5);
\draw[thick,->] (0,0) -- (5, 0);
\node at (5.2,-.1) {\footnotesize$x$};
\node at (-.25,5.2) {\footnotesize$m$};
\draw [thick]   (0,2) to[out=-5,in=110] (4.8,-1);
\node at (5,-1) {\footnotesize$B^s$};
\draw [very thick, gray!60]   (0,3) to[out=-20,in=100] (4,-2);
\node at (4.2,-2) {\footnotesize$B^t$};
\draw[fill=black] (3.95,.4) circle (.075cm);
\node at (3.65,.4) {\footnotesize$y^s$};
\draw[ gray!60, fill= gray!60] (1,2.55) circle (.075cm);
\node at (1.1,2.85) {\footnotesize$y^t$};

\draw[dashed] (4.4,0)--(0,4.35);
\draw[dashed] (0,3.1) -- (5,.4);
\end{tikzpicture}
&
\begin{tikzpicture}

\draw[thick,->] (0,-2) -- (0, 5);
\draw[thick,->] (0,0) -- (5, 0);
\node at (5.2,-.1) {\footnotesize$x$};
\node at (-.25,5.2) {\footnotesize$m$};
\draw [thick]   (0,2) to[out=-5,in=110] (4.8,-1);
\node at (5,-1) {\footnotesize$B^s$};
\draw [very thick, gray!60]   (0,3) to[out=-10,in=100] (4,-2);
\node at (4.2,-2) {\footnotesize$B^t$};
\draw[fill=black] (3.2,1.05) circle (.075cm);
\node at (3.2,.75) {\footnotesize$y^s$};
\draw[dashed] (0,2.9) -- (4.9,.1);

\draw[fill=black] (.5,2.87) circle (.075cm);
\node at (.5,3.2) {\footnotesize$y^t$};
\draw[dashed] (0,3.05) -- (5,1.5);

\draw[thick, dashed] (3.2,1.05) -- (3.2,2);
\draw[decorate,decoration={brace,mirror}, thick] (3.25,1.05) -- (3.25,2);
\node at (4.85,1.55) {\footnotesize $m^t_{\nabla}(x^s)-m^s_{\nabla}(x^s)$};

\draw[thick, dashed] (.5,2.87) -- (.5,2.6);
\draw[decorate,decoration={brace}, thick] (.55,2.87) -- (.55,2.6);
\draw[->] (2,3.5) -- (.67,2.75);
\node at (2.25,3.7) {\footnotesize $m^t_{\nabla}(x^t)-m^s_{\nabla}(x^t)$};

\end{tikzpicture}
\\
(a) Nonconcave utility
&
(b) Nonconcave QL utility
\end{tabular}
\caption{How restrictive is the assumption of concavity of $u$?}
\label{fig:ConcaveViolation}
\end{figure}

Figure \ref{fig:ConcaveViolation} illustrates how restrictive the assumption of concavity of the utility function is when budgets are nonlinear.
We use nonlinear budgets in this case because, as proven by \cite{afriat1967}, the concavity of the utility function has no empirical content when budgets are linear.
Figure \ref{fig:ConcaveViolation}(a) shows that a consumption experiment can satisfy cyclical monotonicity, i.e., be QL rationalizable, but fail GARP for the linearized version of the budgets \citep[see][]{matzkin1991axioms}.
\black{
To proceed with this example we need to make two additional assumptions.
We assume that $\mu^t(x)$ is differentiable and concave.
Then, assuming that preferences are convex (without assuming quasilinearity) implies bundles worse than $y^t$ are below the supporting hyperplane given by $\nabla\mu^t(x^t)$.
Linearized budgets are defined as follows:
$$
\mu^t_{\nabla}(x) = \mu^t(x^t) + \nabla \mu^t(x^t) (x-x^t),
$$
where $\nabla \mu^t(x^t)$ is the gradient of $\mu^t(x)$ evaluated at $x^t$.
By assuming $\mu^t(x)$ to be concave we ensure that $\mu^t(x) \le \mu^t_{\nabla}(x)$, and then passing QL in the linearized experiment implies passing QL in the underlying non-linear experiment.
}
Figure \ref{fig:ConcaveViolation}(b) shows that a consumption experiment can satisfy cyclical monotonicity, i.e., be QL rationalizable, but fail to have a \textit{concave} QL rationalization.
Cyclical monotonicity fails in the linearized budgets in the example since $\mu^t_{\nabla}(x^s)-\mu^s_{\nabla}(x^s)>\mu^t_{\nabla}(x^t)-\mu^s_{\nabla}(x^t)$,
where $\mu^t_{\nabla}$ is the amount of $m$ computed using the gradient of the border of the budget \black{evaluated at the chosen point} as the fixed price.
Hence, we can provide the following result in spirit of \cite{forges2009}.

\begin{rem}
\label{rem:ConcaveRationalization}
    A consumption experiment $E=((x^t,m^t),B^t)_{t=1}^T$ can be rationalized by a concave QL {utility} function if and only if corresponding linearized experiment satisfies cyclical monotonicity.
\end{rem}

In sum, our test significantly generalizes previous results.
Before proceeding, we make a couple of remarks regarding existing results that do not deal with quasilinearity directly.
Quasilinear preferences are a special case of additively separable preferences, and revealed preference theory has addressed separability since \cite{Fishburn1990} seminal result.
The work of \cite{varian1983nonparametric} considers the concave case of additive separability and exploits standard \cite{afriat1967} techniques in order to obtain a test for it.
Varian's test relies crucially on the concavity of each separable component of utility.\footnote{
In this sense the quasilinear analog of the \cite{varian1983nonparametric} is the test developed by \cite{brown2007}.
    }
Recent work of \cite{polisson2018lattice} relaxes this assumption and provides a test for additive separability without concavity.
However, this test requires each separable component to be one-dimensional albeit unknown.\footnote{
        The crucial idea behind the requirement that each component is one-dimensional is that the corresponding set is fully ordered.
        Hence, we treat these notions as synonyms throughout the discussion.
    }
Moreover, \cite{echenique2014testing} shows that the general test for separability of preferences cannot be computationally feasible, while the test we provide is.
We allow the first separable component to be multidimensional, while the second component has to be known and one-dimensional.
Hence, while quasilinearity is a specific case of additive separability, the test we develop is not a special case of any existing tests for separable preferences.
  
Another relevant theory is the augmented utility (AU) introduced by \cite{deb2018revealed}, who develop the Generalized Axiom of Revealed Price Preferences (GAPP).
Augmented utility is close to quasilinearity in the sense that agents have preferences over the monetary costs of consumption goods. GAPP imposes acyclicity (as in GARP) over the revealed price preference relation.
Thus, AU is (in general) not a theory nested within LNU, while QL is nested within both these theories.
However, given that the space of alternatives includes expenditure, then notions of LNU and AU coincide.
Our results relate to GAPP insofar QL is a restriction to LNU in an extended space of alternatives.

\section{Experimental Design}
\label{sec:ExperimentalDesign}
To test for QL, we implement a labor supply experiment with a real effort task: data entry. \black{Our design bears resemblance with recent experimental research that uses real-effort tasks to study time-preferences \citep[][]{augenblick2015working,augenblickrabin2019} and self-control \cite{toussaert2018eliciting}. More recent research has augmented this experimental paradigm to elicit willingness to pay for future tasks \cite[e.g.,][]{carreraetal2022}. Our experiment significantly extends this paradigm by allowing nonlinear incentives. This allows to test for additional features of preferences that are germane to this literature as well as the literature on labor supply in general \citep[see e.g.][]{cherchye2012married,cherchye2017household}. Relatedly, there is a long tradition in applied welfare economics which exploits nonlinearities in budget sets, quasi-linearity and convexity to estimate labor supply responses to taxation \citep[see][]{kleven2016bunching}. Our experimental provides the ideal setting to tests these maintained assumptions. Our design not only closely follows our theoretical results but it is implemented in a highly relevant setting.}

In the experiment, subjects have to choose one option in each of 20 different choice sets. Each choice set has ten separate contracts that specify the number of data entry tasks ($x$) to complete and payment if the tasks are completed ($m$). One of the 20 sets of contracts is randomly selected to secure incentive compatibility, and the subject's choice for that menu is implemented. Figure \ref{fig:BudgetConstruction} provides a graphical representation of the budgets sets faced by subjects. 
    Panel (a) shows a budget set generated by a non-linear wage schedule. 
    The discretized budget is represented as the piece-wise linear budget.
    In this context, the number of tasks (or effort) is a ``bad.''
    We use the slope between $x$ and $x+1$ as the marginal price at $x$ tasks, except for the last case we use the slope between $x-1$ and $x$.
    Panel (b) shows a transformed budget that mimics Figure \ref{fig:GettingM}.
    Denote by $K$ the maximum number of tasks offered to the subject, then $K-x$ is a ``good.''
    By keeping $K$ constant throughout the experiment we maintain all the necessary assumptions about budget set needed to test for QL preferences.

\begin{center}
\begin{figure}[htb]
\centering
\begin{tabular}{cc}
\begin{tikzpicture}
\draw[->,thick] (-.1,0) -- (5,0) node[right] {$x$};
\draw[->,thick] (0,-.1) -- (0,5) node[above] {$m$};


\draw[gray,dashed, thick](0,0) -- (.5,1.5*.5^.5) -- (1,1.5*1^.5) -- (2,1.5*2^.5) -- (2.5,1.5*2.5^.5) -- (3,1.5*3^.5) -- (3.5,1.5*3.5^.5) -- (4,1.5*4^.5) -- (4,0);

\draw[fill=black] (.5,1.5*.5^.5) circle (.075cm);
\draw[fill=black] (1,1.5*1^.5) circle (.075cm);
\draw[fill=black] (1.5,1.5*1.5^.5) circle (.075cm);
\draw[fill=black] (2,1.5*2^.5) circle (.075cm);
\draw[fill=black] (2.5,1.5*2.5^.5) circle (.075cm);
\draw[fill=black] (3,1.5*3^.5) circle (.075cm);
\draw[fill=black] (3.5,1.5*3.5^.5) circle (.075cm);
\draw[fill=black] (4,1.5*4^.5) circle (.075cm);

\draw[->, thick, gray] (2.2,1) -> (1.2,3);

\end{tikzpicture}
&
\begin{tikzpicture}
\draw[->,thick] (-.1,0) -- (5,0) node[right] {$K-x$};
\draw[->,thick] (0,-.1) -- (0,5) node[above] {$m$};


\draw[gray,dashed, thick](4-0,0) -- (4-.5,1.5*.5^.5) -- (4-1,1.5*1^.5) -- (4-2,1.5*2^.5) -- (4-2.5,1.5*2.5^.5) -- (4-3,1.5*3^.5) -- (4-3.5,1.5*3.5^.5) -- (4-4,1.5*4^.5) -- (4-4,0);

\draw[fill=black] (4-.5,1.5*.5^.5) circle (.075cm);
\draw[fill=black] (4-1,1.5*1^.5) circle (.075cm);
\draw[fill=black] (4-1.5,1.5*1.5^.5) circle (.075cm);
\draw[fill=black] (4-2,1.5*2^.5) circle (.075cm);
\draw[fill=black] (4-2.5,1.5*2.5^.5) circle (.075cm);
\draw[fill=black] (4-3,1.5*3^.5) circle (.075cm);
\draw[fill=black] (4-3.5,1.5*3.5^.5) circle (.075cm);
\draw[fill=black] (4-4,1.5*4^.5) circle (.075cm);

\draw[->, thick, gray] (1.1,1) -> (2.1,3);

\end{tikzpicture}
\\
(a) Initial Budget Set
&
(b) Reflected Budget Set
\end{tabular}
\caption{Construction of the Budget}
\label{fig:BudgetConstruction}
\end{figure}
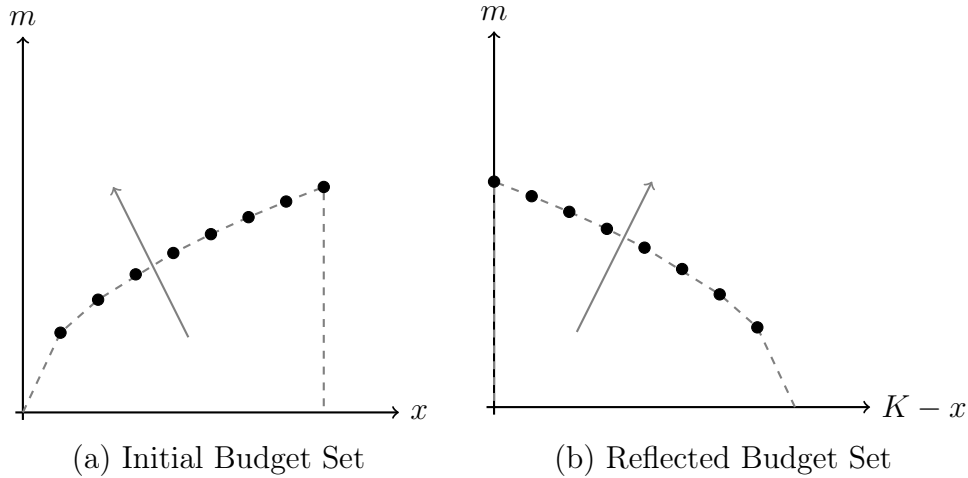
\end{center}

\noindent

\begin{figure}[p]
\begin{subfigure}[a]{1\linewidth}
\centering
\begin{tikzpicture}
\begin{axis}[grid=major,
            xlabel=number of tasks, 
            ylabel=wage (\$),
            xmin=.8,
            xmax = 10.2,
            ymin = 0,
            ymax = 30,
            scale=1.2,
            xtick={1,2,3,...,10}
            ]

\addplot[mark=diamond*, dashed] table [y=b1, x=tasks]{Budgets.dat};
\addplot[mark=square*, dotted] table [y=b2, x=tasks]{Budgets.dat};
\addplot[mark=diamond*, loosely dotted] table [y=b3, x=tasks]{Budgets.dat};
\addplot[mark=triangle*, densely dotted] table [y=b4, x=tasks]{Budgets.dat};
\addplot[mark=diamond*, dashdotted] table [y=b5, x=tasks]{Budgets.dat};
\addplot[mark=diamond*, dashed] table [y=b6, x=tasks]{Budgets.dat};
\addplot[mark=triangle*, dotted] table [y=b7, x=tasks]{Budgets.dat};
\addplot[mark=o, gray] table [y=b8, x=tasks]{Budgets.dat};
\addplot[mark=square, gray] table [y=b9, x=tasks]{Budgets.dat};
\addplot[mark=diamond, gray] table [y=b10, x=tasks]{Budgets.dat};
\addplot[mark=triangle, gray] table [y=b11, x=tasks]{Budgets.dat};
\addplot[mark=otimes, gray] table [y=b12, x=tasks]{Budgets.dat};
\addplot[mark=diamond*, dotted] table [y=b13, x=tasks]{Budgets.dat};
\addplot[mark=triangle*, dashed] table [y=b14, x=tasks]{Budgets.dat};
\addplot[mark=diamond*, dashdotted] table [y=b15, x=tasks]{Budgets.dat};
\addplot[mark=diamond*, dotted] table [y=b16, x=tasks]{Budgets.dat};
\addplot[mark=diamond*, loosely dotted] table [y=b17, x=tasks]{Budgets.dat};
\addplot[mark=otimes*, dotted] table [y=b18, x=tasks]{Budgets.dat};
\addplot[mark=triangle*, dashed] table [y=b19, x=tasks]{Budgets.dat};
\addplot[mark=triangle*, loosely dotted] table [y=b20, x=tasks]{Budgets.dat};
\end{axis}
\end{tikzpicture}
\caption{Budgets used in the experiment}
\label{fig:RealBudgets}
\end{subfigure}\vfill
\begin{subfigure}[b]{1\linewidth}
\centering
\includegraphics[scale=.3]{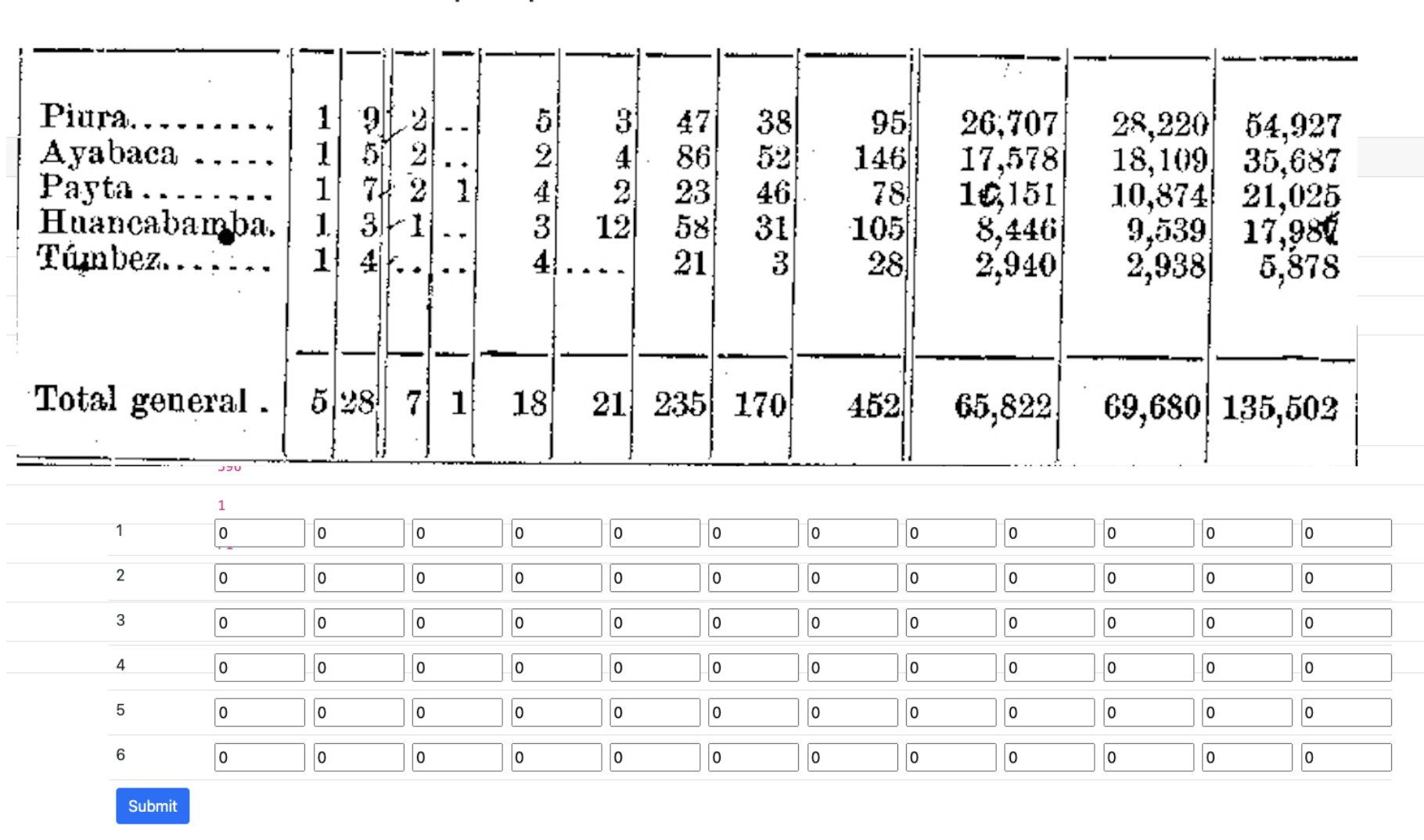}
\caption{Data entry task Interface}
\label{fig:TaskInterface}
\end{subfigure}\vfill
\begin{subfigure}[c]{1\linewidth}
\centering
\includegraphics[scale=.3]{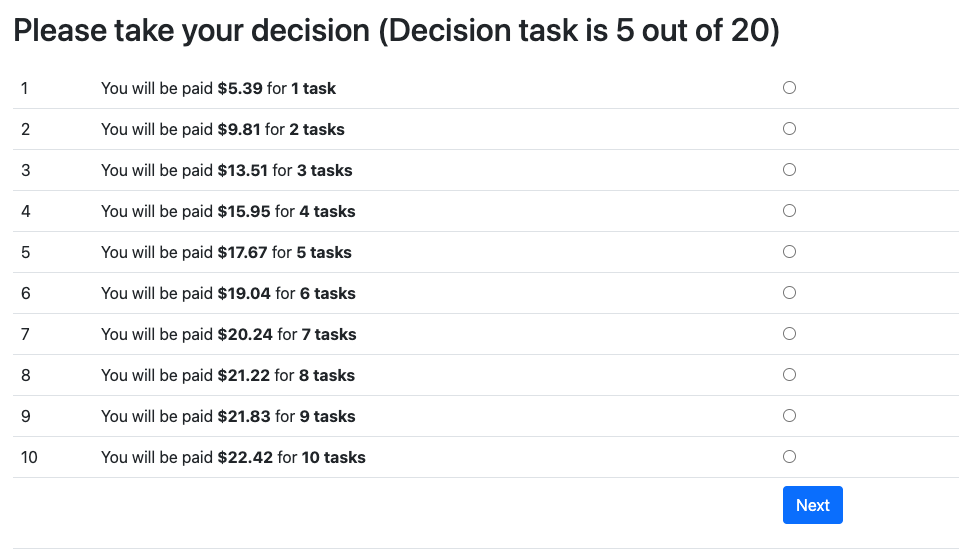}
\caption{Decision task Interface}
\label{fig:DecisionInterface}
\end{subfigure}\vfill
\caption{Implementation of the experiment}
\label{fig:ExperimentalDetails}
\end{figure}

Figure \ref{fig:RealBudgets} depicts the actual menus used in the experiment. The minimal wage used is \$3 (for 1 task), and the maximal wage is about \$27 (for ten tasks). 
Figure \ref{fig:TaskInterface} presents a typical data entry task and the interface used by subjects. The table is an extract from the 1876 Peruvian population census. 
In the real effort tasks, subjects were asked to enter the data from the Peruvian Census in a data-entry grid (see Figure \ref{fig:TaskInterface}).
Each cell in the data-entry grid corresponds to a cell in the scanned table from the census. To make sure that subjects entered the data correctly, the interface used control sums.
The last row in each census table, called ``Total general,'' is the total of the entries above each column and is never equal to zero.
Similarly, the numbers in the last column equal the sum of the numbers in two preceding columns. We use this to create control sums.
If a subject submitted entries that did not match the control sums, she was told that the data entered was in error, and was informed which row and/or column had a problem.
To make the task a real-effort job, a task was deemed complete, or a contract fulfilled, once errors were absent.
Figure \ref{fig:DecisionInterface} shows the interface given to subjects to choose among contracts. All choice sets contain ten contracts, and contracts were always ordered from 1 task to 10 tasks. 

The timeline of the experiment is as follows. First, to allow subjects to form an expectation of the time it takes to complete an average task, we implement a practice data entry task in which consistency checks are enforced. Second, subjects choose contracts from each one of the 20 budgets. Finally, one of the 20 budgets is chosen at random, and the chosen contract is implemented. All the tasks (census extracts) are randomly selected from a pool of pre-selected sections of comparable length and readability. The experiment was implemented in September-October 2021 using oTree \citep{chen2016} at the Texas A\&M University (TAMU) experimental lab. A total of 65 undergraduate students participated in the experiment with average earnings of \$15.

Table \ref{tab:Descriptive} presents basic summary statistics of the experiment. Subjects choose 7.03 (s.d. 2.75) tasks to complete on average. The number of choices varies across menus. The mean and median spread of the number of tasks is 4.75 and 5. Eleven out of 65 subjects always chose the same number of tasks. Decisions of individuals who do not vary their labor supply across menus can always be rationalized.

\begin{table}[htb]
\centering
\begin{tabular}{l|c}
\hline\hline
Mean Choice                  & 7.03 \\
Average Standard Deviation   & 2.75 \\ \hline
Min Spread                   & 0    \\
Max Spread                   & 9    \\
Mean Spread                  & 4.75 \\
Median Spread                  & 5.00 \\
Mode Spread                  & 5.00 \\
Standard Deviation of Spread & 3.01 \\ \hline\hline
\end{tabular}
\caption{Descriptive Statistics of Choices}
\label{tab:Descriptive}
\end{table}

\section{Experimental Results}
\label{sec:ExperimentalResults}
This section presents the results of the experiment. 
We denote the concave version of locally nonsatiated utility (LNU) as C-LNU, and the concave version of quasilinear utility (QLU) as C-QLU. 
We illustrate the main results. 
First, we show that concavity and quasilinearity are equally restrictive. 
That is, C-LNU and QLU have similar empirical support.
Second, the joint assumption of concavity and quasilinearity (C-QLU) is more restrictive than either concavity alone (C-LNU) or quasilinearity alone (QLU).
Third, subjects consistent with C-QLU exhibit behavioral patterns that are significantly different from subjects consistent with LNU/QLU/C-LNU.
In particular, subjects consistent with C-QLU exhibit significantly less elastic labor supply.

We find that 11 out of 65 subjects are consistent with all the theories considered.
Note that the subjects consistent with all four theories are the only subjects consistent with at least one theory.
These are the subjects who chose the same number of tasks in all sets of contracts. To allow for decision errors, we evaluate consistency with different hypotheses on preferences using the \cite{houtman1985determining} Index (HMI). The HMI is the maximum percent of choices consistent with a theory (in our case, LNU, QLU, C-LNU or C-QLU). In our data, a perfectly consistent subject has an HMI of one, and the lowest possible HMI is 0.05 (1/20). Since inconsistent choices are counted as errors, the HMI provides an upper bound of the probability of deliberate choice.\footnote{
    \cite{dembo2019rationality} shows that this estimate is biased while the unbiased estimator is not computationally tractable.
}
We use the HMI because it is the simplest to calculate and interpret for all of the theories we consider. For robustness, Appendix B shows that similar results hold if we use the CCEI.

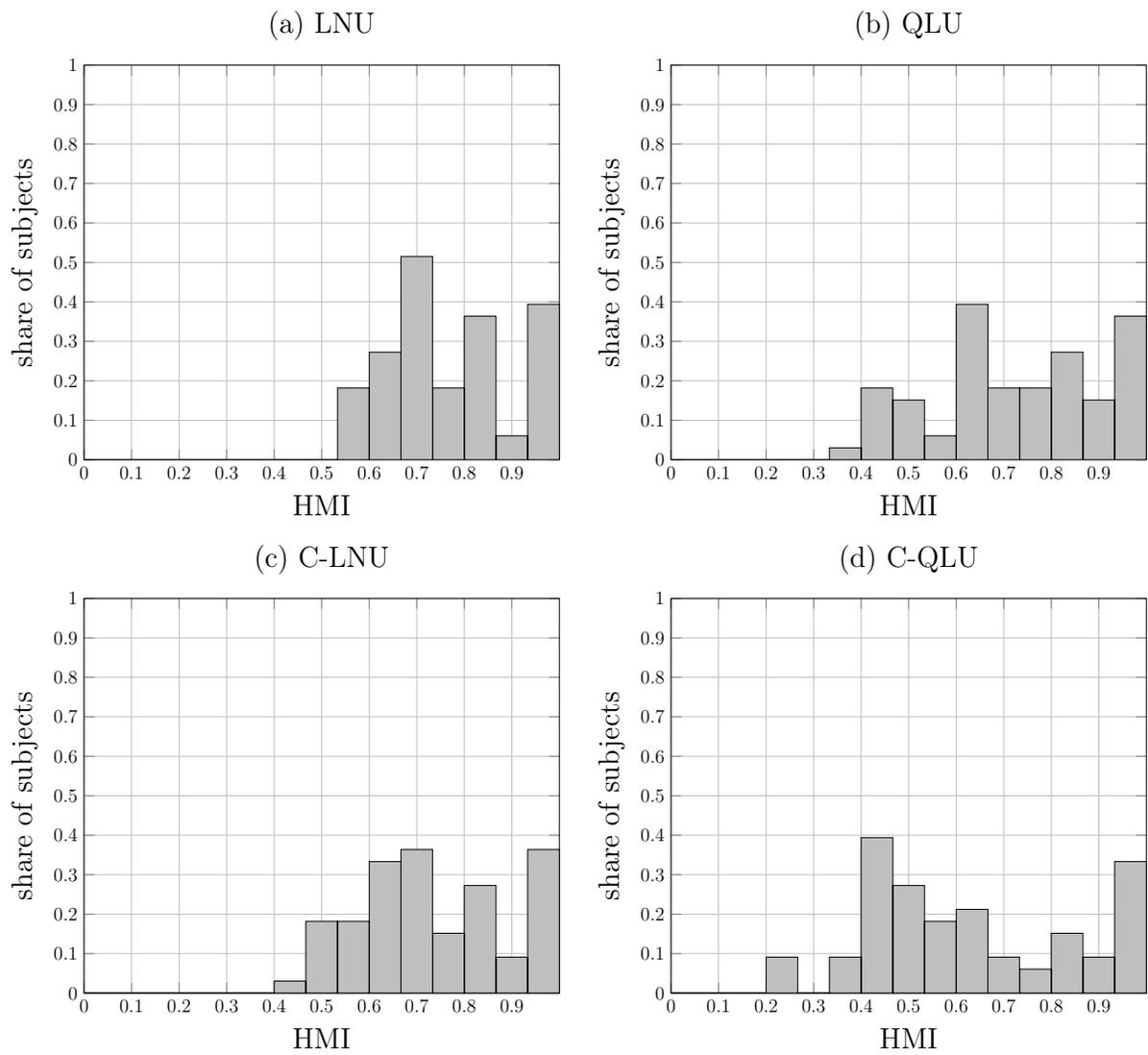
\begin{figure}[htp]
\centering
\resizebox{\linewidth}{!}{
\begin{tabular}{cc}
\begin{tikzpicture}
    \begin{axis}[
    ymin =0, 
    ymax=1,
    xmax=1,
    xmin=0,
    grid=major,
    xlabel = HMI,
    ylabel = share of subjects,
    yticklabel style={
        /pgf/number format/fixed,
        /pgf/number format/precision=5
    },
    scaled y ticks=false,
    ytick = {0,.1,.2,.3,.4,.5,.6,.7,.8,.9,1},
    xtick = {0,.1,.2,.3,...,1},
    title = (a) LNU,
    ]
    \addplot [
        hist={bins=15, data min = 0, data max = 1},
        fill=gray!50,
        y filter/.expression={y/33},
    ] 
    table [y=real_LNU_HMI]{HMI_Results.dat};
    \end{axis}
    \end{tikzpicture}
    &
    \begin{tikzpicture}
    \begin{axis}[
    ymin =0, 
    ymax=1,
    xmax=1,
    xmin=0,
    grid=major,
    xlabel = HMI,
    ylabel = share of subjects,
    yticklabel style={
        /pgf/number format/fixed,
        /pgf/number format/precision=5
    },
    scaled y ticks=false,
    ytick = {0,.1,.2,.3,.4,.5,.6,.7,.8,.9,1},
    xtick = {0,.1,.2,.3,...,1},
    title = (b) QLU,
    ]
    \addplot [
        hist={bins=15, data min = 0, data max = 1},
        fill=gray!50,
        y filter/.expression={y/33},
    ] 
    table [y=real_QLU_HMI]{HMI_Results.dat};
    \end{axis}
    \end{tikzpicture}
    \\
    \begin{tikzpicture}
    \begin{axis}[
    ymin =0, 
    ymax=1,
    xmax=1,
    xmin=0,
    grid=major,
    xlabel = HMI,
    ylabel = share of subjects,
    yticklabel style={
        /pgf/number format/fixed,
        /pgf/number format/precision=5
    },
    scaled y ticks=false,
    ytick = {0,.1,.2,.3,.4,.5,.6,.7,.8,.9,1},
    xtick = {0,.1,.2,.3,...,1},
    title = (c) C-LNU,
    ]
    \addplot [
        hist={bins=15, data min = 0, data max = 1},
        fill=gray!50,
        y filter/.expression={y/33},
    ] 
    table [y=real_concave_LNU_HMI]{HMI_Results.dat};
    \end{axis}
    \end{tikzpicture}
    &
    \begin{tikzpicture}
    \begin{axis}[
    ymin =0, 
    ymax=1, 
    xmax=1,
    xmin=0,
    grid=major,
    xlabel = HMI,
    ylabel = share of subjects,
    ytick = {0,.1,.2,.3,.4,.5,.6,.7,.8,.9,1},
    xtick = {0,.1,.2,.3,...,1},
    yticklabel style={
        /pgf/number format/fixed,
        /pgf/number format/precision=5
    },
    scaled y ticks=false,
    title = (d) C-QLU,
    ]
    \addplot [
        hist={bins=15, data min = 0, data max = 1},
        fill=gray!50,
        y filter/.expression={y/33},
    ]
    table [y=real_concave_QLU_HMI]{HMI_Results.dat};

    \end{axis}
    \end{tikzpicture}
\end{tabular}
}
\caption{Distribution of HMIs}
\label{fig:HMIhists}
\end{figure}

Figure \ref{fig:HMIhists} presents the distributions of the HMI for all of the theories.
Panels (a), (b), (c), and (d) show the distributions for LNU, QLU, C-LNU, and C-QLU, respectively.
The mean HMI for LNU is 0.77 (SD 0.14), the mean HMI for QLU is 0.73 (SD 0.19), the mean HMI for C-LNU is 0.74 (SD 0.17), and the mean HMI for C-QLU is 0.63 (SD 0.23). The mean (medians) of C-LNU, QLU, and C-QLU are significantly lower than for LNU.\footnote{
    Differences in both means and medians are significant at $p<.01$.} 
This is not surprising since they are restrictions on LNU. QLU and C-LNU perform similarly in rationalizing the behavior in the experiments. Finally, the mean (median) HMI for C-QLU is significantly lower than that of QLU and C-LNU.\footnote{
    Differences in both means and medians are significant at $p<.01$.
} Imposing convexity to QL is costly.

\subsection{Validating the Utility Hypotheses}
Proper tests of alternative utility hypotheses require accounting for false positives or the possibility that a subject passes a theory by chance. 
We start by constructing a test for LNU.
The null hypothesis for this case is random behavior, and the alternative hypothesis is consistency with LNU.
The critical value for this hypothesis is a pre-determined level of the HMI (see below).
To estimate the distribution of the parameter under the null hypothesis, we calculate the HMI of 400,000 randomly generated subjects.
In particular, for each menu of each random subject, we select a contract according the empirical distribution of choices in the menu.\footnote{
    For a more detailed discussion of the method used to generate random choices and conducting hypothesis testing, see \cite{cherchye2020permutation}.
}
We treat the distribution of HMI resulting from this procedure as the asymptotic distribution of HMI under the null hypothesis.
The significance level $\alpha$ is calculated using the $100-\alpha$ percentile of the distribution of HMIs of random subjects. We use a one-side test given that the HMI is a distance to rationality index.
We consider experimental subjects whose HMI is above this threshold as being consistent with LNU since the null hypothesis is rejected at the given significance level.

Next, to account for the nestedness of QLU and C-LNU theories into LNU.
Conditional pass rates are calculated using the subpopulation of random subjects passing LNU at the $\alpha$ significance level.
This provides a test of restrictions on LNU, as now the null hypothesis is (random) behavior consistent with LNU.
We consider two ways to generate cutoffs for C-QLU. One uses the subpopulation of random subjects consistent with C-LNU, and the other uses the subpopulation of random subjects consistent with QLU.\footnote{
    Of the 400000 random subjects, 20000 are consistent with LNU. These are used to construct cutoffs for QLU and C-LNU.  Finally, 1,000 of random subjects consistent with QLU or C-LNU are used to generate cutoffs for C-QLU.
    }

\smallskip

\begin{table}[htb]
\begin{tabular}{l|ccc|ccc}
\hline
         & \multicolumn{3}{c|}{Pass Rates}                                                                                                                                                                                      & \multicolumn{3}{c}{HMI cutoffs} \\ \cline{2-7} 
         & $p=.10$                                                               & $p=.05$                                                              & $p=.01$                                                               & $p=.10$   & $p=.05$  & $p=.01$  \\ \hline
LNU      & \begin{tabular}[c]{@{}c@{}}77\% (50/65) \\ $[.65, .87]$\end{tabular}  & \begin{tabular}[c]{@{}c@{}}77\% (50/65) \\ $[.65, .87]$\end{tabular} & \begin{tabular}[c]{@{}c@{}}51\% (33/65) \\ $[.38, .63]$\end{tabular}  & .70       & .70      & .75      \\ \hline
QLU      & \begin{tabular}[c]{@{}c@{}}86\% (43/50)  \\ $[.73, .94]$\end{tabular} & \begin{tabular}[c]{@{}c@{}}86\% (43/50) \\ $[.73, .94]$\end{tabular} & \begin{tabular}[c]{@{}c@{}}79\% (26/33) \\ $[.61, .91]$\end{tabular}  & .65       & .65      & .75      \\
C-LNU    & \begin{tabular}[c]{@{}c@{}}82\% (41/50) \\ $[.69, .91]$\end{tabular}  & \begin{tabular}[c]{@{}c@{}}82\% (41/50) \\ $[.69, .91]$\end{tabular} & \begin{tabular}[c]{@{}c@{}}73\% (24/33)  \\ $[.55, .87]$\end{tabular} & .70       & .70      & .80      \\ \hline
C-QLU(C) & \begin{tabular}[c]{@{}c@{}}69\% (28/41) \\ $[.52, .82]$\end{tabular}  & \begin{tabular}[c]{@{}c@{}}59\% (24/41) \\ $[.42, .74]$\end{tabular} & \begin{tabular}[c]{@{}c@{}}79\% (19/24)\\ $[.58, .93]$\end{tabular}   & .65       & .70      & .80      \\
C-QLU(Q) & \begin{tabular}[c]{@{}c@{}}56\% (24/43) \\ $[.40, .71]$\end{tabular}  & \begin{tabular}[c]{@{}c@{}}56\% (24/43) \\ $[.40,.71]$\end{tabular}  & \begin{tabular}[c]{@{}c@{}}73\% (19/26) \\  $[.52, .88]$\end{tabular} & .70       & .70      & .80      \\ \hline
\end{tabular}
\caption{Left panel: pass rates with absolute numbers in parentheses and confidence intervals for the pass rates in brackets. Tests for QLU and C-LNU are conditional on passing LNU; tests for C-QLU are conditional on either QLU or C-LNU. Right panel: HMI cutoffs for the given ``significance'' level.}
\label{tab:PassRates}
\end{table}

Table \ref{tab:PassRates} presents pass rates (left panel) and thresholds (right panel) for the theories considered.
The percentages are the pass rates, which are unconditional for LNU and conditional for the remaining. The numbers in parentheses are the subjects who pass a theory and their comparison group. The numbers in brackets are the 95\% confidence intervals of the pass rate calculated using the Clopper-Pearson procedure.\footnote{
    The pass rate is a standard Binomial variable, where ``success'' is rejecting the null hypothesis (subject passed) and ``failure'' is not rejecting the null (subject failed). The Clopper-Pearson procedure provides exact confidence intervals. 
    Results are similar if we estimate confidence intervals using the CLT (see Table \ref{tab:PassRatesCLT}).
The significance levels we consider are $p\in\{.10,.05,.01\}$ corresponding to the 90\%, 95\%, and 99\% cutoffs. The right panel presents the HMI thresholds used in the analysis.}

Table \ref{tab:PassRates} shows support for LNU, C-LNU, and QLU. Depending on the tolerance level of the test, the pass rate for LNU is between 51\% and 77\%. The pass rate for QLU is between 79\% and 86\%, and the pass rate for C-LNU is between 73\% and 82\%.
The support for QLU and C-LNU holds even if we consider unconditional pass rates, which are more demanding for nested theories. The unconditional pass rate for QLU is between 40\% and 66\% and for C-LNU is between 37\% and 63\%. Next, we consider adding a concavity restriction to QL (denoted C-QLU(Q)) or to C-LNU (denoted C-QLU(C)). The conditional pass rates decrease significantly after imposing this restriction ($p<.01$) for thresholds $p=.10$ and $p=.05$. The unconditional pass rates for C-QLU(C) are 43\% ($[.31, .56]$), 37\% ($[.25,.50]$), and 29\% ($[.19,.42]$) for $p=.10$, $p=.05$, and $p=.01$ and for C-QLU(Q) are 37\% ($[.25,.50]$), 37\% ($[.25,.50]$), and 29\% ($[.19,.42]$) for $p=.10$, $p=.05$, and $p=.01$. Full results on unconditional pass rates are provided in Appendix B.

\subsection{Individual Choices}

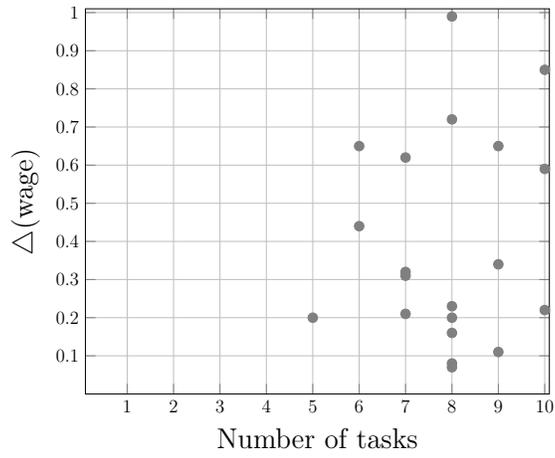
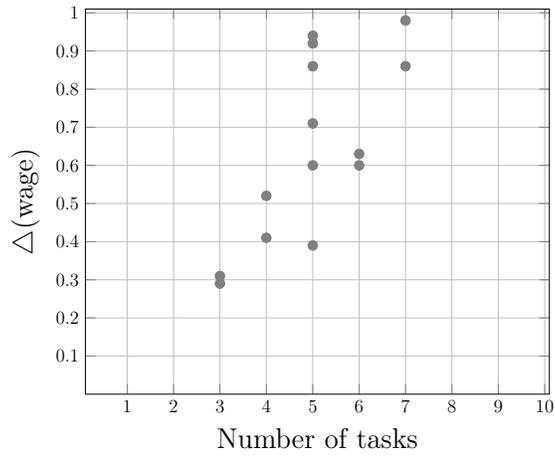
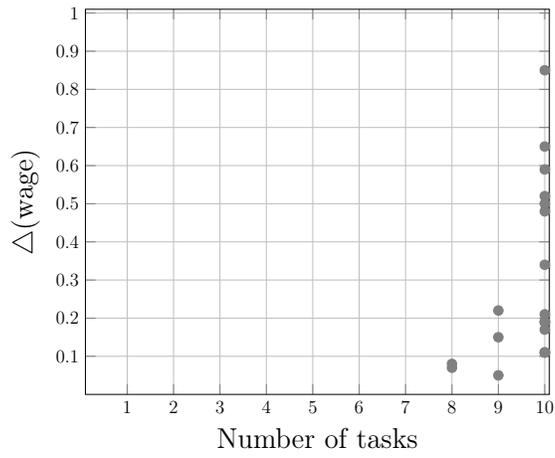
\begin{figure}[htp]
    \centering
    \begin{subfigure}[t]{1\linewidth}
    \centering
    \scalebox{0.9}{
    \begin{tikzpicture}
     \begin{axis}[
    ymin =0, 
    ymax=1.01,
    xmax=10.1,
    xmin=0.1,
    grid=major,
    xlabel = Number of tasks,
    ylabel = $\triangle$(wage),
    scaled y ticks=false,
    ytick = {.1,.2,.3,.4,.5,.6,.7,.8,.9,1},
    xtick = {1,2,3,...,10},
    ]
   \addplot[only marks, mark=otimes*, gray] table [x=choice, y=wage]{notQLU_subject33.dat};
    \end{axis}
    \end{tikzpicture}
    }
    \caption{LNU but not QLU (subject 33)}
    \label{fig:noQLU}
    \end{subfigure}
    
    \vspace{.5cm}
    
    \begin{subfigure}[t]{1\linewidth}
    \centering
    \scalebox{0.9}{
    \begin{tikzpicture}
     \begin{axis}[
    ymin =0, 
    ymax=1.01,
    xmax=10.1,
    xmin=0.1,
    grid=major,
    xlabel = Number of tasks,
    ylabel = $\triangle$(wage),
    scaled y ticks=false,
    ytick = {.1,.2,.3,.4,.5,.6,.7,.8,.9,1},
    xtick = {1,2,3,...,10},
    ]
   \addplot[only marks, mark=otimes*, gray] table [x=choice, y=wage]{QLU_subject9.dat};
    \end{axis}
    \end{tikzpicture}
    }
    \caption{QLU (subject 9)}
    \label{fig:QLU}
    \end{subfigure}
    
    \vspace{.5cm}
    
    \begin{subfigure}[t]{1\linewidth}
    \centering
    \scalebox{0.9}{
    \begin{tikzpicture}
     \begin{axis}[
    ymin =0, 
    ymax=1.01,
    xmax=10.1,
    xmin=0.1,
    grid=major,
    xlabel = Number of tasks,
    ylabel = $\triangle$(wage),
    scaled y ticks=false,
    ytick = {.1,.2,.3,.4,.5,.6,.7,.8,.9,1},
    xtick = {1,2,3,...,10},
    ]
   \addplot[only marks, mark=otimes*, gray] table [x=choice, y=wage]{CQLU_subject7.dat};
    \end{axis}
    \end{tikzpicture}
    }
    \caption{C-QLU (subject 7)}
    \label{fig:CQLU}
    \end{subfigure}
    \caption{Examples of the choices of subjects}
    \label{fig:SupplyCurves}
\end{figure}

This section presents individual-level results. 
Because quasilinearity is frequently used as an assumption that holds only locally, i.e. for relatively small changes in labor supply or changes due to minor changes in incentives, we will make clear when the assumption is assumed to be local or global, i.e. applicable to the entire domain of labor supply. 
We start by showing supply functions for different theories.
To gain intuition, recall that in the case QLU, the optimal choice of tasks depends only on the slope of the budget constraint (e.g., in the differentiable case, for optimal choice $x^*$ we have that $u'(x_i^*)=m'(x_i^*)$ or $x_i^*=(u')^{-1}(m'(x_i^*))$). 
As mention in the previous section, we approximate the slope of budget constraint by the change in the wage per task completed. 
Figure \ref{fig:SupplyCurves} shows scatter plots of the change in wages as a function of the number of tasks chosen. 
We consider 3 subjects \black{that illustrate well different behavioral patterns}:
\begin{itemize}
    \item [] Subject 33 who is consistent with LNU but not QLU;
    \item [] Subject 9 who is consistent with QLU;
    \item [] Subject 7 who is consistent with C-QLU.
\end{itemize}

Due to noisy behavior (HMI $<1$), the scatter plots include choices that would need to be excluded to satisfy preference restrictions exactly. Subject 33, depicted on Figure \ref{fig:noQLU}, is consistent with LNU but not QL. As the graph shows, there is no clear relationship between choice of tasks and changes in wages due to income effects.
Subject 9, depicted in Figure \ref{fig:QLU}, is consistent with QLU. Since QLU implies a constant marginal disutility of effort, we expect a clustering of changes in wages at each level of effort.\footnote{
Note that subjects are choosing over discretized budgets and therefore marginal conditions are only approximations.
This pattern is observed for levels 3, 4, 6, and 7 tasks, but not at level 5.
Concavity further requires that the changes in wages increase in the number of tasks (the marginal disutility of effort is increasing).
That is, the maximal $\triangle$(wage) faced while choosing $x$ tasks should be below the minimal $\triangle$(wage) faced while choosing $y$ tasks for $y>x$. Subject 7, depicted in Figure \ref{fig:CQLU}, provides an example of this. The implication is satisfied for $\triangle$(wage)  corresponding to choices of 8 and 10 tasks. However, none of the choices of 9 tasks can be rationalized exactly by C-QLU.}

\begin{figure}[h]
    \centering
\begin{subfigure}[t]{.45\linewidth}
\begin{tikzpicture}
\begin{axis}[grid=major,
            xlabel=HMI, 
            ylabel=Mean Spread,
            xmin=0,
            xmax = 1,
            ymin = 0,
            ymax = 5,
            scale=.9,
            xtick = {0,.1,.2,...,1},
            ytick = {0,.5,1,...,5},
            legend pos=south west,
            legend style={font=\small},
            ]
\addplot[mark=diamond] table [y=qlu_spread, x=hmi]{Spread.dat};
\addlegendentry{QLU}

\addplot[mark=square*, dashed] table [y=concave_qlu_spread, x=hmi]{Spread.dat};
\addlegendentry{C-QLU}

\addplot[mark=otimes*, dashed, gray] table [y=lnu_spread, x=hmi]{Spread.dat};
\addlegendentry{LNU}

\addplot[mark=triangle*, dashed, gray] table [y=concave_lnu_spread, x=hmi]{Spread.dat};
\addlegendentry{C-LNU}
\end{axis}
\end{tikzpicture}
\caption{Spread}
\label{fig:Spread}
\end{subfigure}
\hfill
\begin{subfigure}[t]{.45\linewidth}
\begin{tikzpicture}
\begin{axis}[grid=major,
            xlabel=HMI, 
            ylabel=Mean Variance,
            xmin=0,
            xmax = 1,
            ymin = 0,
            ymax = 3,
            scale=.9,
            xtick = {0,.1,.2,...,1},
            ytick = {0,.5,1,...,3},
            legend pos=south west,
            legend style={font=\small},
            ]
\addplot[mark=diamond] table [y=qlu_spread, x=hmi]{Variance.dat};
\addlegendentry{QLU}

\addplot[mark=square*, dashed] table [y=concave_qlu_spread, x=hmi]{Variance.dat};
\addlegendentry{C-QLU}

\addplot[mark=otimes*, dashed, gray] table [y=lnu_spread, x=hmi]{Variance.dat};
\addlegendentry{LNU}

\addplot[mark=triangle*, dashed, gray] table [y=concave_lnu_spread, x=hmi]{Variance.dat};
\addlegendentry{C-LNU}
\end{axis}
\end{tikzpicture}
\caption{Variance}
\label{fig:Variance}
\end{subfigure}
\caption{{Dispersion analysis of choice by HMI}}
\label{fig:SpreadAndElasticity}
\end{figure}
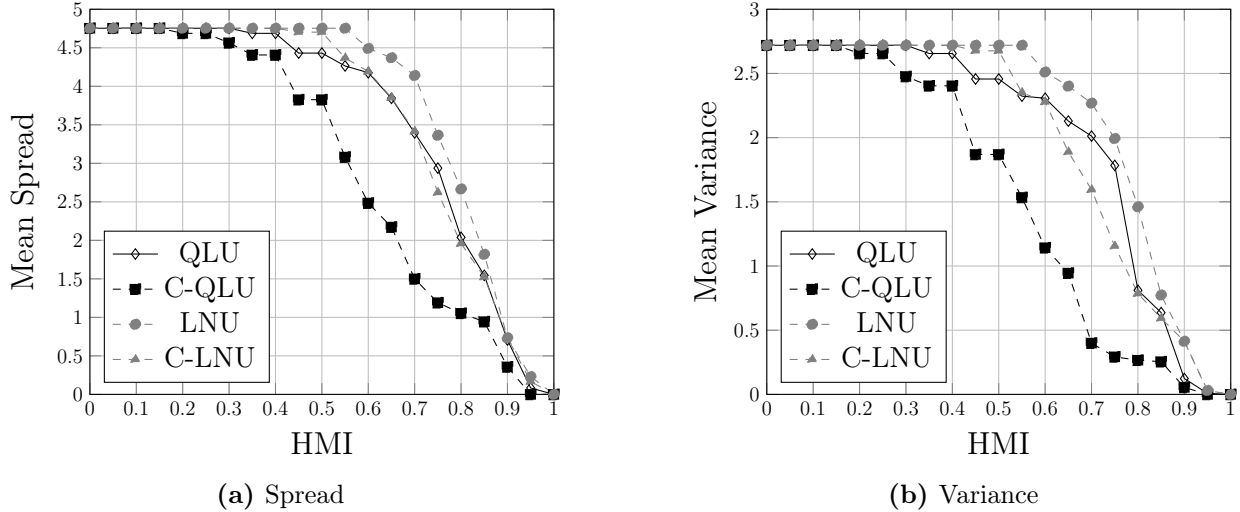

Finally, to visualize how consistency with different behavioral assumption relates with labor supply elasticity, we analyze the dispersion of choices by HMI level. We focus on two indicators: spread and variance. That is, for every subject we compute the spread (difference between maximum and minimal) and variance of the number of tasks chosen.
Figure \ref{fig:SpreadAndElasticity} plots the average of these measures for each HMI level and each behavioral assumption. For instance, Figure \ref{fig:Spread} shows that for subjects with an HMI$=.6$ for LNU the average spread is 4.5.
Figure \ref{fig:Variance} shows that for subjects with an HMI$=.6$ for LNU the average variance is 2.5.
The mean spread for LNU is the largest regardless the HMI level.
Mean spreads for QLU and C-LNU are comparable to LNU and each other.
The mean spread for C-QLU is the smallest.
Results based on the variance are consistent with those using spreads.

\section{Concluding Remarks}
We provide necessary and sufficient conditions for a set of observed choices to be rationalizable  by quasilinear (QL) {preferences}. 
These conditions apply to choices over compact and downward closed budget sets and do not require the utility function to be concave.
The test of QL that we present applies to a large class of problems. First, it applies to consumer problems in the presence of distortions due to taxes, subsidies, or nonlinear pricing. 
Second, it can be extended to test for QL preferences in strategic situations when QL is invoked (e.g., auction theory). 
It can also be used to analyze labor supply decisions, where nonlinear prices (wages for efforts or output) appear naturally. {Finally, we test the assumption of QL preferences using a real-effort task experiment. This is not only the most straightforward environment in which the theory can be tested but also reproduces the labor-income decision setting in which QL is commonly assumed. Our main empirical result is that QL tests that do not allow for the possibility of non-convex preferences significantly underestimate the support for QL preferences.}

The empirical support of QL in our experiment might appear to be at odds with recent research on income effects \citep[see e.g.][]{cesarini2017effect,giupponi2019income,golosov2021americans}. These studies, however, consider large changes in income (lotteries or significant subsidies), while our laboratory setting deals with small stakes (less than \$30). QL might still be a reasonable assumption when only small changes in income are considered. The generality of our test for QL preferences allows us to expand the contexts in which this assumption can be tested. For instance, \cite{strack2021dynamic} show that while time inconsistency cannot be identified from preference reversals alone, it can be identified if preferences are quasilinear. This is so because, under quasilinearity, a cardinal measure of preference reversals can be obtained. The same applies to contexts in which willingness to accept is used to assess the consistency of behavior. Our work provides a way to test this maintained assumption.

\bibliographystyle{model2-names}
\bibliography{refs}

\clearpage

\appendix
\section{Proof of Theorem \ref{WeakRepresentation}}
\begin{proof}
{\bf ($\Rightarrow$)}
Consider a function $v(x,m) = u(x) +m$ that rationalizes the consumption experiment.
{
For clarity, we maintain the notation $m^t(x^s)$, $m^t(x^t)$ and $m^t(x)$ for the rest of the proof.
}
Then, for every sequence of observations $k_1,k_2,\ldots, k_n \in \{1,\ldots,T\}$, the following is true:
$$
u(x^{k_{j+1}})+m^{k_{j+1}}(x^{k_{j+1}})\geq u(x^{k_{j}}) + m^{k_{j+1}}(x^{k_{j}}), 
$$
\noindent where $k_{n+1}=k_1$.
We can simplify this and obtain the following inequalities:
\begin{equation*}
\begin{cases}
& m^{k_1}(x^{k_1}) - m^{k_1}(x^{k_2}) \geq  u(x^{k_2})  - u(x^{k_1}), \\
& m^{k_2}(x^{k_2}) - m^{k_2}(x^{k_3}) \geq  u(x^{k_3})  - u(x^{k_2}), \\
&\ldots \ \ldots \  \ldots \  \ldots  \  \ldots \  \ldots \  \ldots \  \ldots \  \ldots \  \ldots \\
& m^{k_n}(x^{k_n}) - m^{k_n}(x^{k_1}) \geq  u(x^{k_1})-u(x^{k_n}).
\end{cases}    
\end{equation*}
Summing up these inequalities, we obtain the following:

$$
m^{k_1}(x^{k_1}) - m^{k_1}(x^{k_2})+ m^{k_2}(x^{k_2}) - m^{k_2}(x^{k_3}) +\ldots+ m^{k_n}(x^{k_n}) - m^{k_n}(x^{k_1})\geq 0.
$$
This is exactly the cyclical monotonicity condition.
\newline 

\noindent
{\bf ($\Leftarrow$)}
This part of the proof is split in two parts.
First, we construct the subutility numbers corresponding to the observed choices and show that the observed chosen point $(x^t,m^t)$ is at least as good as any other point $(x^s,m^t(x^s))$.
Next, we extend the subutility function to the entire $\R^n_+$ and show that corresponding QLU rationalizes the data.

Let 
$$
u^t = \min\{m^{k_1}(x^{k_1})-m^{k_1}(x^{t})+\ldots+m^{k_{n}}(x^{k_n})-m^{k_n}(x^{k_{n-1}})\}
$$
over all sequences in the data, including those with repeating elements.\footnote{
    In terms of graph theory, this is equivalent to searching for the shortest walk (on a weighted directed graph), rather than a shortest path (on a weighted directed graph). However, the shortest walk is not always well-defined. A sufficient condition for it to be well-defined is the absence of negative cycles, which is guaranteed by cyclical monotonicity if we define the complete directed graph with vertexes as observations and weights as $w_{s\rightarrow t} = m^t(x^t)-m^t(x^s)$.
}
Note that all the sequences are starting with $k_1$, however, the real ``anchoring'' element of the sequence is $k_2=t$ for the observation $t\in T$.
We will show that the minimum is well-defined.
It will be enough to show that there will be no cycles in the minimal sequence, since the rest will follow from the fact that the minimum is taken over finite sums of finite numbers. 
So assume, to the contrary, that the minimum sequence contains a cycle; then we have 
$$
\ldots+ \left( m^{k_n}(x^{k_n})-m^{k_n}(x^{k_{n-1}}) + \ldots + m^{k_1}(x^{k_1}) - m^{k_1}(x^{k_n})\right) + \ldots.
$$
However, cyclical monotonicity implies that this term is greater or equal than zero, and hence, excluding it would make the sequence even smaller.
That contradicts the original assumption that the sequence was the smallest.

Further, we show that such a construction of $u^t$ will guarantee that the following system of inequalities is satisfied.
For any $t,s\in \{1,\ldots, T\}$ we want to show that
$$
u^t - u^s \geq m^t(x^s)-m^t(x^t).
$$
By the construction of $u(x)$, we can guarantee that 
$$
u^s \leq m^t(x^t)-m^t(x^s)+u^t,
$$
since 
$$
u^t = m^{k_1}(x^{k_1})-m^{k_1}(x^t)+\ldots+m^{k_{n}}(x^{k_n})-m^{k_n}(x^{k_{n-1}})
$$
for some (minimal) sequence, and we construct $u(x^s)$ using that minimal sequence.
Recall that we allowed taking every sequence, including those with repeating elements, so we can add any element to the existing sequence and the utility level $u(x^s)$ will not exceed the value of the new extended sequence.
Therefore,
$$
u^t-u^s\geq u^t - \left(m^t(x^t)-m^t(x^s)+u(x^t)\right)=m^t(x^s)-m^t(x^t).
$$
This concludes the first part of the proof.
Next, we use the numbers constructed above to recover the entire utility function.

For every $x\in X$, let
$$
u(x) = \min\limits_{t\in \{1,\ldots, T\}} \{u^t+m^t(x^t)-m^t(x)\}.
$$
Note that since $m^t$ is continuous and monotone, so is $u(x)$.

First, we will show that for every $t\in \{1,\ldots, T\}$, $u(x^t)=u^t$.
As we have shown above for every $s\in \{1,\ldots, T\}$ for which $m^s(x^t)$ is defined,
$$
u^t\leq u^s + m^s(x^s)-m^s(x^t).
$$
Therefore, $u(x^t)=u^t$. Next, we show that the constructed utility rationalizes the data, that is $v(x,m)\leq v(x^t,m^t(x^t))$ for every $(x,m)\in B^t$, where $v(x,m) = u(x)+m$.
By the construction of $u(x)$, we know that $u(x) = \min\limits_{t\in \{1,\ldots, T\}} \{u^t+m^t(x^t)-m^t(x)\} \leq u^t+m^t(x^t)-m^t(x)$.
Therefore, $u(x)+m \leq u(x)+m^t(x) \leq u(x^t)+m^t(x^t)$.
\end{proof}

Given the proof of the Theorem \ref{WeakRepresentation} one can immediately infer the following remark.

\begin{rem}
\label{rem:LinearProgrammingConditions}
A consumption experiment $E=((x^t,m^t),B^t)_{t=1}^T$ can be rationalized by a QLU function if and only if there is a monotone function $u: C\rightarrow \R$ such that
    $$
    u(x^t)-u(x^s) \geq m^t(x^s)-m^t(x^t) \text{ for every } t,s\in \{1,\ldots,T\}.
    $$
where $C = \bigcup\limits_{t\in \{1,\ldots,T\} } \{x^t\}$ is the set of chosen points.
\end{rem}

Remark \ref{rem:LinearProgrammingConditions} crucially important for the following reason.
Cyclical monotonicity is an elegant condition, it may be less computationally tractable especially once mixed with measures of distance to rationality.
However, the linear programming approach provides the definitely computationally tractable way of testing QLU rationalizability.


\setcounter{figure}{0}
\renewcommand{\thefigure}{C.\arabic{figure}}
\setcounter{table}{0}
\renewcommand{\thetable}{C.\arabic{table}}
\setcounter{section}{2}
\clearpage
\section{Additional Empirical Results}
{This appendix presents additional empirical analysis.
We first present additional results using the HMI index. We then present analysis using the Critical Cost Efficiency Index (CCEI) introduced by \cite{afriat1973}. The results are robust to alternative measures of distance to rationality.}

\subsection{Additional Results for HMI}
\label{app:Robustness}
{We present two additional tables. Table \ref{tab:UnconditionalRates} presents unconditional pass rates. Table \ref{tab:PassRatesCLT} presents an alternative construction of confidence intervals using normal approximations instead of the exact procedure.}

\begin{table}[htb]
\begin{tabular}{l|ccc|ccc}
\hline
         & \multicolumn{3}{c|}{Pass Rates}                                                                                                                                                                                      & \multicolumn{3}{c}{HMI cutoffs} \\ \cline{2-7} 
         & $p=.10$                                                               & $p=.05$                                                              & $p=.01$                                                               & $p=.10$   & $p=.05$  & $p=.01$  \\ \hline
LNU      & \begin{tabular}[c]{@{}c@{}}77\% (50/65) \\ $[.65, .87]$\end{tabular}  & \begin{tabular}[c]{@{}c@{}}77\% (50/65) \\ $[.65, .87]$\end{tabular} & \begin{tabular}[c]{@{}c@{}}51\% (33/65) \\ $[.38, .63]$\end{tabular}  & .70       & .70      & .75      \\ \hline
QLU      & \begin{tabular}[c]{@{}c@{}}66\% (43/65)  \\ $[.54, .74]$\end{tabular} & \begin{tabular}[c]{@{}c@{}}66\% (43/65) \\ $[.53, .74]$\end{tabular} & \begin{tabular}[c]{@{}c@{}}40\% (26/65) \\ $[.38, .53]$\end{tabular}  & .65       & .65      & .75      \\
C-LNU    & \begin{tabular}[c]{@{}c@{}}63\% (41/65) \\ $[.50, .75]$\end{tabular}  & \begin{tabular}[c]{@{}c@{}}63\% (41/65) \\ $[.50, .75]$\end{tabular} & \begin{tabular}[c]{@{}c@{}}37\% (24/65)  \\ $[.25, .50]$\end{tabular} & .70       & .70      & .80      \\ \hline
C-QLU(C) & \begin{tabular}[c]{@{}c@{}}43\% (28/65) \\ $[.31, .56]$\end{tabular}  & \begin{tabular}[c]{@{}c@{}}37\% (24/65) \\ $[.25, .50]$\end{tabular} & \begin{tabular}[c]{@{}c@{}}29\% (19/65)\\ $[.19, .42]$\end{tabular}   & .65       & .70      & .80      \\
C-QLU(Q) & \begin{tabular}[c]{@{}c@{}}37\% (24/65) \\ $[.25, .50]$\end{tabular}  & \begin{tabular}[c]{@{}c@{}}37\% (24/65) \\ $[.25,.50]$\end{tabular}  & \begin{tabular}[c]{@{}c@{}}29\% (19/65) \\  $[.19, .42]$\end{tabular} & .70       & .70      & .80      \\ \hline
\end{tabular}
\caption{
Left panel: Pass rates for each theory considered. Unlike Table \ref{tab:PassRates} in the main text, this table reports unconditional pass rates.
The first number is the percentage pass rate; the fraction in parentheses is the numbers of subjects used to compute the pass rates; the numbers brackets are the 95\% confidence intervals.
Right panel: HMI cutoffs for different $p$-levels.
}
\label{tab:UnconditionalRates}
\end{table}

{Table \ref{tab:UnconditionalRates} replicates the results of Table \ref{tab:PassRates} \black{but uses unconditional pass rates instead}.
These numbers serve for purely illustrative purposes, as it is not totally correct to analyze nested theories using unconditional pass rates.
Recall that the nested structure can be summarized as follows.}
$$
\text{C-QLU(C)} \subseteq \text{C-LNU} \subseteq \text{ LNU } 
\text{ and }
\text{C-QLU(Q)} \subseteq \text{QLU} \subseteq \text{ LNU } 
$$
{That is, C-QLU is nested within both C-LNU and QLU, while both C-LNU and QLU are nested within LNU.
Finally, C-LNU and QLU are independent theories.}

\begin{table}[htb]
\begin{tabular}{l|ccc|ccc}
\hline
         & \multicolumn{3}{c|}{Pass Rates}                                                                                                                                                                                    & \multicolumn{3}{c}{HMI cutoffs} \\ \cline{2-7} 
         & $p=.10$                                                              & $p=.05$                                                              & $p=.01$                                                              & $p=.10$   & $p=.05$  & $p=.01$  \\ \hline
LNU      & \begin{tabular}[c]{@{}c@{}}77\% (50/65) \\ $[.67, .87]$\end{tabular} & \begin{tabular}[c]{@{}c@{}}77\% (50/65) \\ $[.67, .87]$\end{tabular} & \begin{tabular}[c]{@{}c@{}}51\% (33/65) \\ $[.39, .63]$\end{tabular} & .70       & .70      & .75      \\ \hline
QLU      & \begin{tabular}[c]{@{}c@{}}86\% (43/50) \\ $[.76, .96]$\end{tabular} & \begin{tabular}[c]{@{}c@{}}86\% (43/50)\\  $[.76, .96]$\end{tabular} & \begin{tabular}[c]{@{}c@{}}79\% (26/33)\\ $[.65, .93]$\end{tabular}  & .65       & .65      & .75      \\
C-LNU    & \begin{tabular}[c]{@{}c@{}}82\% (41/50) \\ $[.71, .93]$\end{tabular} & \begin{tabular}[c]{@{}c@{}}82\% (41/50)\\ $[.71, .93]$\end{tabular}  & \begin{tabular}[c]{@{}c@{}}73\% (24/33)\\ $[.57, .88]$\end{tabular}  & .70       & .70      & .80      \\ \hline
C-QLU(C) & \begin{tabular}[c]{@{}c@{}}69\% (28/41)\\  $[.54, .83]$\end{tabular} & \begin{tabular}[c]{@{}c@{}}59\% (24/41) \\ $[.43, .74]$\end{tabular} & \begin{tabular}[c]{@{}c@{}}79\% (19/24)\\ $[.63, .96]$\end{tabular}  & .65       & .70      & .80      \\
C-QLU(Q) & \begin{tabular}[c]{@{}c@{}}56\% (24/43)\\  $[.41, .71]$\end{tabular} & \begin{tabular}[c]{@{}c@{}}56\% (24/43)\\ $[.41,.71]$\end{tabular}   & \begin{tabular}[c]{@{}c@{}}73\% (19/26)\\ $[.56, .91]$\end{tabular}  & .70       & .70      & .80      \\ \hline
\end{tabular}
\caption{
Left panel: Pass rates for each theory considered. Unlike Table \ref{tab:PassRates} in the main text, this table reports confidence intervals for pass rates using CLT approximations.
The first number is the percentage pass rate; the fraction in parentheses is the numbers of subjects used to compute the pass rates; the numbers brackets are the 95\% confidence intervals.
Right panel: HMI cutoffs for different $p$-levels.
}
\label{tab:PassRatesCLT}
\end{table}

{Table \ref{tab:PassRatesCLT} replicates the results of Table \ref{tab:PassRates} using the CLT approximations for the confidence intervals of pass rates.
Note that unlike the confidence intervals generated by Clopper-Pearson procedure (exact confidence intervals), the CLT confidence intervals are symmetric.
Thus, they might run out of the domain: returning values above one or below zero.
Even though there are no bound violations in our data we prefer to report the exact confidence intervals.
Using the CLT confidence intervals does not alter the results presented in Table \ref{tab:PassRates}.}

\subsection{Analysis Using CCEI}
{Another measure of distance to rationality is the Critical Cost Efficiency Index (CCEI) introduced by \cite{afriat1973}.
The definition of the CCEI in the augmented consumption space we consider is, however, ambiguous.
An interpretation of the CCEI is the share of wealth to be left on the table needed to make an agent rational.
This can be done in linear budget sets straightforwardly.
Assume all prices ($p$) are set up such that for each budget set income is normalized to be equal to $1$.
Then $p^t y \le 1$ defines the ``revealed preference relation.''
The CCEI redefines this constraint by introducing $e\in[0,1]$ such that $p^t y \le e$ defines a new revealed preference relation.
In this case $1-e$ provides the relative measure of the wealth one needs to sacrifice in order to rationalize observed choices.
Further, the CCEI finds the maximal $e$ (minimizing potential wealth losses) such that observed choices are rationalizable.}

\begin{figure}[htb]
    \centering
\begin{tabular}{cc}
\begin{tikzpicture}
    \draw [thick, ->] (0,-.1) -> (0,5) node[left] {$m$};
    \draw [thick, ->] (-.1,0) -> (5,0) node[below] {$x$};
    \draw[dashed] (2,2) -- (2,0);
    \draw[dashed] (2*.8,2*.8) -- (2*.8,0);
    
    \draw[thick] (4.25,0) -- (2,2) -- (0,2.75);
    \draw node[right] at (2.1,2.1) {$g(y)=1$};
    \draw[->] (2*.95,2*.95) -> (2*.85,2*.85);
    \draw node[left] at (2,1) {$g(y)=e$};
    \draw[dashed] (4.25*.8,0) -- (2*.8,2*.8) -- (0,2.75*.8);
\end{tikzpicture}
     & 
\begin{tikzpicture}
    \draw [thick, ->] (0,-.1) -> (0,5) node[left] {$m$};
    \draw [thick, ->] (-.1,0) -> (5,0) node[below] {$x$};
    
    \draw[dashed] (2,2) -- (2,0);
    \draw[thick] (4.25,0) -- (2,2) -- (0,2.75);
    \draw node[right] at (2.1,2.1) {$g(y)=1$};
    \draw[->] (1.9,2*.95) -> (1.9,2*.85);
    \draw node[left] at (2,1.5) {$g(y)=e$};
    \draw[dashed] (4.25,0) -- (2,2*.8) -- (0,2.75*.8);
\end{tikzpicture}
\\
(a) Just Noticeable Differences CCEI
&
(b) Wealth CCEI
\end{tabular}
    \caption{Introducing CCEI in Augmented Space}
    \label{fig:CCEI}
\end{figure}
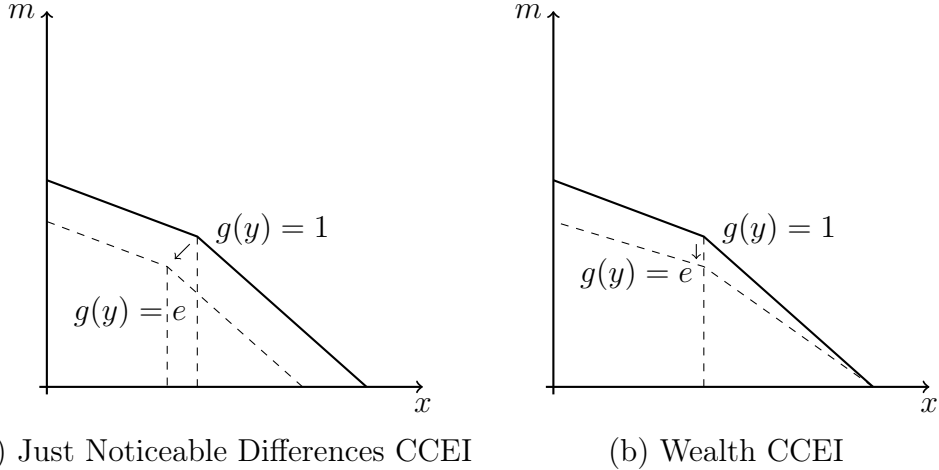

{However, in the augmented consumption space, ``wealth'' (wage in our context) is already one of the dimensions of the consumption space.
Thus one option is to consider only changes in wages.
This corresponds to the case described on the Figure \ref{fig:CCEI}(b).
An alternative explanation of the CCEI is based on the idea of ``thick indifference curves'' or ``just noticeable differences'' as axiomatized by \cite{dziewulski2020just}.
This interpretation allows us to disregard the fact that the consumption space includes money already.
This interpretation allows shifting the budget downwards as presented in Figure \ref{fig:CCEI}(a).
The nonlinearity of the budget set, however, produces another complication.
Since we are using piece-wise linear budgets, it matters if the value of $x$ at which price regime changes Varies with the level of the CCEI.
The just noticeable difference interpretation of the CCEI (see Figure \ref{fig:CCEI}(a)) implies that the ``break-point'' varies with $e$.
The wealth interpretation of the CCEI (see Figure \ref{fig:CCEI}(b)) does not allow the break-point to vary with $e$.
Next we present the results for the pass rates for both interpretations of CCEI.}

\subsubsection{CCEI based on Wealth}

\begin{table}[htb]
\begin{tabular}{l|ccc|ccc}
\hline
         & \multicolumn{3}{c|}{Pass Rates}                                                                                                                                                                                    & \multicolumn{3}{c}{CCEI cutoffs} \\ \cline{2-7} 
         & $p=.10$                                                              & $p=.05$                                                              & $p=.01$                                                              & $p=.10$   & $p=.05$   & $p=.01$  \\ \hline
LNU      & \begin{tabular}[c]{@{}c@{}}83\% (54/65) \\ $[.72, .91]$\end{tabular} & \begin{tabular}[c]{@{}c@{}}75\% (49/65) \\ $[.63, .85]$\end{tabular} & \begin{tabular}[c]{@{}c@{}}52\% (34/65) \\ $[.40, .65]$\end{tabular} & .96       & .98       & 1        \\ \hline
QLU      & \begin{tabular}[c]{@{}c@{}}91\% (49/54) \\ $[.80, .97]$\end{tabular} & \begin{tabular}[c]{@{}c@{}}96\% (47/49)\\  $[.86, .99]$\end{tabular} & \begin{tabular}[c]{@{}c@{}}79\% (27/34)\\ $[.62, .91]$\end{tabular}  & .90       & .92       & .96      \\
C-LNU    & \begin{tabular}[c]{@{}c@{}}83\% (45/54) \\ $[.71, .92]$\end{tabular} & \begin{tabular}[c]{@{}c@{}}82\% (40/49)\\ $[.68, .91]$\end{tabular}  & \begin{tabular}[c]{@{}c@{}}79\% (27/34)\\ $[.62, .91]$\end{tabular}  & .90       & .94       & .99      \\ \hline
C-QLU(C) & \begin{tabular}[c]{@{}c@{}}91\% (41/45)\\  $[.79, .98]$\end{tabular} & \begin{tabular}[c]{@{}c@{}}85\% (34/40) \\ $[.70, .94]$\end{tabular} & \begin{tabular}[c]{@{}c@{}}78\% (21/27)\\ $[.58, .91]$\end{tabular}  & .88       & .92       & .96      \\
C-QLU(Q) & \begin{tabular}[c]{@{}c@{}}84\% (41/54)\\  $[.70, .92]$\end{tabular} & \begin{tabular}[c]{@{}c@{}}72\% (34/47)\\ $[.57,.84]$\end{tabular}   & \begin{tabular}[c]{@{}c@{}}78\% (21/27)\\ $[.58, .91]$\end{tabular}  & .88       & .92       & .96      \\ \hline
\end{tabular}
\caption{Left panel: pass rates with absolute numbers in parentheses and confidence intervals for the pass rates in brackets. Tests for QLU and C-LNU are conditional on passing LNU; tests for CQLU are conditional on either QLU or C-LNU. Right panel: CCEI cutoffs for the given ``significance'' level.}
\label{tab:PassRatesCCEI}
\end{table}

Table \ref{tab:PassRatesCCEI} presents the results as in Table \ref{tab:PassRates} using the CCEI based on changes in wealth only (version from Figure \ref{fig:CCEI}(b)) as the measure of distance to rationality.
The pass rates for LNU and QLU are comparable to those using HMI.
However, comparing C-LNU to QLU we observe that the pass rates for C-LNU are lower. This is evidence that the assumption of concavity is more restrictive than quasilinearity.
Overall we see that the pass rates for C-QLU(Q) are lower than those for QLU. We observe that concavity is a restrictive assumption. However, pass rates for C-QLU(C) and C-LNU are comparable.
Thus, adding the quasilinearity assumption to {C-LNU} does not seem to be restrictive.These results are consistent with the results presented in Table \ref{tab:PassRates}.

\subsubsection{CCEI based on Just Noticeable Differences}
\begin{table}[htb]
\begin{tabular}{l|ccc|ccc}
\hline
         & \multicolumn{3}{c|}{Pass Rates}                                                                                                                                                                                    & \multicolumn{3}{c}{CCEI cutoffs} \\ \cline{2-7} 
         & $p=.10$                                                              & $p=.05$                                                              & $p=.01$                                                              & $p=.10$   & $p=.05$   & $p=.01$  \\ \hline
LNU      & \begin{tabular}[c]{@{}c@{}}78\% (51/65) \\ $[.67, .88]$\end{tabular} & \begin{tabular}[c]{@{}c@{}}68\% (44/65) \\ $[.55, .79]$\end{tabular} & \begin{tabular}[c]{@{}c@{}}63\% (41/65) \\ $[.50, .75]$\end{tabular} & .99       & .99       & 1        \\ \hline
QLU      & \begin{tabular}[c]{@{}c@{}}92\% (47/51) \\ $[.81, .98]$\end{tabular} & \begin{tabular}[c]{@{}c@{}}93\% (41/44)\\  $[.81, .99]$\end{tabular} & \begin{tabular}[c]{@{}c@{}}85\% (35/41)\\ $[.71, .94]$\end{tabular}  & .94       & .96       & .97      \\
C-LNU    & \begin{tabular}[c]{@{}c@{}}86\% (44/51) \\ $[.74, .94]$\end{tabular} & \begin{tabular}[c]{@{}c@{}}84\% (37/44)\\ $[.70, .93]$\end{tabular}  & \begin{tabular}[c]{@{}c@{}}68\% (28/41)\\ $[.52, .82]$\end{tabular}  & .86       & .93       & .99      \\ \hline
C-QLU(C) & \begin{tabular}[c]{@{}c@{}}95\% (42/44)\\  $[.85, .97]$\end{tabular} & \begin{tabular}[c]{@{}c@{}}89\% (33/37) \\ $[.75, .97]$\end{tabular} & \begin{tabular}[c]{@{}c@{}}93\% (26/28)\\ $[.77, .99]$\end{tabular}  & .92       & .92       & .96      \\
C-QLU(Q) & \begin{tabular}[c]{@{}c@{}}89\% (42/47)\\  $[.77, .96]$\end{tabular} & \begin{tabular}[c]{@{}c@{}}80\% (33/41)\\ $[.65,.91]$\end{tabular}   & \begin{tabular}[c]{@{}c@{}}77\% (27/35)\\ $[.60, .90]$\end{tabular}  & .86       & .92       & .96      \\ \hline
\end{tabular}
\caption{Left panel: pass rates with absolute numbers in parentheses and corresponding confidence intervals in brackets. Tests for QLU and C-LNU are conditional on passing LNU; tests for CQLU are conditional on either QLU or C-LNU. Right panel: CCEI cutoffs for the given ``significance'' level.}
\label{tab:PassRatesCCEI2}
\end{table}

Table \ref{tab:PassRatesCCEI2} presents results using the just noticeable differences interpretation of the CCEI (version from Figure \ref{fig:CCEI}(a)).
We see that the main results still hold.
First, we see that pass rates for QLU condition on LNU are high.
Moreover, using this version of the CCEI we obtain higher pass rates for QLU than C-LNU.
We also see that pass rates for C-LNU are weakly lower than those of C-QLU(C) for all significance levels.
This means that under the assumption of concavity adding quasilinearity is costless.
Finally, we see that pass rates for C-QLU(Q) are lower than those of QLU. 
This agrees, accounting for noise, with the results obtained with alternative measures of rationality. We conclude that our main finding of quasilinearity is not more restrictive than concavity (see Table \ref{tab:PassRates}) does not depend on the measure of distance to rationality utilized.

\clearpage
\section{{Computing HMI (Online Only)}}
\label{sec:mipcodes}

We use mixed integer programming (MIP) to compute the HMI. 
MIP allows us to use the \say{big-M} method to implement the conditional logic.
To compute the HMI, we need to find the maximal rationalizable subset of the data.
That is, we need to exclude the minimal amount of observations possible.
For every observation $t\in T$, we define $v_t \in \R_+$, the utility value for the chosen point (an arbitrary value).
Given that we have finite amount of observation, we ensure that $v_t$ is bounded.
We denote the upper bound of $v_t$ by $V$.
Let $\delta_t \in \lbrace 0,1 \rbrace$ be an indicator of whether observation $t\in T$ is excluded or not.
That is, $\delta_t = 1$ if the observation $t\in T$ is excluded, and $0$ otherwise.
In order to exclude the observation we use the \say{big-M} ($M\rightarrow\infty$). For $B^t = \lbrace (x,m) : m\le \mu^t(x) \rbrace$, we have

\begin{equation}
    \label{eq:lnu_mip}
    \begin{cases}
    \sum\limits_{t\in T} \delta_t \rightarrow \min\limits_{\delta_t, v_t} & \\
    v_t + (\delta_t+\delta_s) M \ge v_s & \text{ if } (x^s,m^s) \in B^t \text{ for every } t,s\in T \\
    0 \le V_t \le V &  \text{ for every } t\in T \\
    \end{cases}
\end{equation}

System \eqref{eq:lnu_mip} presents the mixed integer program for LNU.
The first line guarantees that we find the maximal rationalizable subset of data, by minimizing the number of observations to be excluded.
Last line directly imposes the bounds on $v_t\in \R_+$ for every $t\in T$.
The second line implements the test for LNU, one can easily see that excluding $ (\delta_t+\delta_s) M$ term the inequality directly implements GARP.
The term $(\delta_t+\delta_s) M$ guarantees that the inequality is excluded (trivially satisfied) if either observation $t\in T$ is excluded ($\delta_t =1$) or observation $s\in T$ is excluded ($\delta_s =1$).

Next introduce necessary definitions for QLU MIP.
Recall that QLU assumes that 
$$
v(x,m) = u(x) + m.
$$
This logic has to be implemented in the corresponding MIP.
Let $u_t \in R$ be a bounded utility values, with explicit upper bound being equal to $U$.
Let $m_{t,s} = \mu^t(x_s)$ that is the monetary value that corresponds to $x^t$ in budget $B^t$.
Note that by construction $m_{t,t} = m^t$.
Finally, both $\delta_t \in \lbrace 0,1\rbrace$ and $M\rightarrow \infty$ keep similar interpretations.

\begin{equation}
    \label{eq:qlu_mip}
    \begin{cases}
    \sum\limits_{t\in T} \delta_t \rightarrow \min\limits_{\delta_t, u_t} & \\
    u_t + m_{tt} + (\delta_t+\delta_s) M \ge u_s + m_{ts} & \text{ for every } t,s\in T \\
    0 \le u_t \le U &  \text{ for every } t\in T \\
    \end{cases}
\end{equation}

System \eqref{eq:qlu_mip} presents the mixed integer program for QLU.
The first line guarantees that we find the maximal rationalizable subset of data, by minimizing the number of observations to be excluded.
Last line directly imposes the bounds on $u_t$ for every $t\in T$.
The second line without the $(\delta_t+\delta_s) M$ implements the cyclical monotonicity condition using linear program.
The connection between the second line and the cyclical monotonicity condition is provided in the proof of Proposition \ref{WeakRepresentation}.
The term $(\delta_t+\delta_s) M$ guarantees that the inequality is excluded (trivially satisfied) if either observation $t\in T$ is excluded ($\delta_t =1$) or observation $s\in T$ is excluded ($\delta_s =1$).

In order to construct the test for C-LNU and C-QLU we can use the systems \eqref{eq:lnu_mip} and \eqref{eq:qlu_mip} correspondingly with a slight modification.
The only difference is that we need to use the linearized budget ($\mu^t_{\nabla}(x)$) instead of the original budget.
Using the linearized budget implies that if there is a solution to the MIP, then there is a concave representation.
Reasoning behind the latter statement is that if the budgets are linear, then concavity of the utility function has no additional empirical content for neither C-LNU nor C-QLU.
In terms of technical implementation we use MATLAB with YALMIP interface package for the linear and mixed integer programming.
More details on YALMIP package can be found  \href{https://yalmip.github.io/tutorials/}{here}.
The solver used is GUROBI, more details can be found \href{https://www.gurobi.com/}{here}.

\clearpage
\section{Experimental Instructions (Online Only)}
\label{sec:experimentprocedures}

The instructions presented below consist of four parts and were presented subject on the screen of their desktops before the corresponding stage.
\begin{enumerate}
    \item First part of the instructions (firsy page), welcomed subjects to the experiment and explained some general rules of behavior in the lab.
    \item Part 1 explains the data entry (real effort) task.
    It also explains the internal consistency checks implemented and explains the potential error messages appearing if data was not inputted correctly.
    Finally, it provides some hints on deciphering some of the hard to read input.
    \item Part 2 explains the decision task and explains that choices made in this stage are binding.
    \item Part 3 reminds of the procedure of selection of the contract to be implemented.
    In addition it reminds subjects of the rules of the data entry tasks.
    Finally, it emphasizes that after completing this task subjects are good to go and do not need to wait for other participants.
\end{enumerate}

You can find the experiment \href{https://quasilinear-job-experiment.herokuapp.com/}{here} and the source code \href{https://github.com/mlfreer/ql_experiment}{here}.



\section*{Supplementary material}

# Offline Instructions

Welcome! Thank you for joining us in this experiment. You have a chance to earn **$5 to $25,** depending on the choices you make and the effort you put in. This is a study of individual decision-making, so your earnings depend on your decisions only. The experiment consists of three tasks, where the instructions for each task will be shown separately before each task begins.

## PART 1

In this part, you will be asked to input data. You will be presented with a scanned version of one of the pages of the 1876 Peruvian Census together with a screen to input the data. Figure 1 shows a screenshot of the actual screen you will face in this part of the experiment. If you do not see the submit button, please scroll to the bottom of the screen.

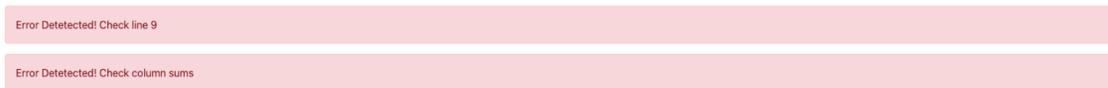

Figure 1. Screenshot of the Task

The system has internal checks to make sure that the data is input correctly. The system will inform you if you have made a typo. The system will also help you locate typos by highlighting the row or column where the typos are. Figure 2 illustrates how the screen identifies the location of a typo.

Figure 2. Potential Error Message

Given the difficulty of the task, the system will allow for a few mistakes. The permitted number of mistakes increases with the number of rows and columns in the table. The computer will enable you to progress to the next task once you have entered the data without exceeding the permitted number of mistakes. To check and fix potential mistakes, please check the following **control sums (using a calculator if necessary)**:

- The number at the bottom of every column is the sum of all the numbers above it.
- In every row, the sum of the numbers in the two columns previous to the last column must add up to the number in the last column.

In addition to the guidelines above, please feel free to employ these **hints**:

- The symbol "**...**" must be replaced with **0**.
- Number **0** could indeed be **6** or **8**.
- Number **9** could indeed be **8**.

Once you have input the data with sufficiently few errors, the computer will allow you to submit your decision and proceed to the following table. This is illustrated in Figure 3.

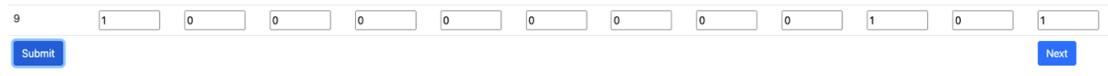

*Figure 3. Successful Data Submission*

**Thank You and Good Luck!**

# PART 2

Thank you for completing Part I of the experiment! We are now proceeding to the next stage of the experiment. Please read the instructions carefully. You will be able to proceed shortly after all participants finish Part I.

In this part, you will be asked to make **20** decisions. Each decision consists of choosing one of 10 possible alternatives from a menu. Each alternative in a menu is a contract. Each contract specifies how much you will be paid if you complete a certain number of data entering tasks. The tasks are similar to those you just completed in Part I. Each menu contains the payment you would receive if you complete 1, 2, 3, …, or 10 tasks. Figure 4 shows an example of a menu you might face. The first option offers $4.02 for completing one task, $6.51 for completing two tasks, etc.

## Please take your decision (Decision task is 1 out of 20)

| | | |
|---|---|---|
| 1 | You will be paid **$4.02** for **1 task** | ○ |
| 2 | You will be paid **$6.51** for **2 tasks** | ○ |
| 3 | You will be paid **$8.09** for **3 tasks** | ○ |
| 4 | You will be paid **$9.35** for **4 tasks** | ○ |
| 5 | You will be paid **$10.34** for **5 tasks** | ○ |
| 6 | You will be paid **$11.0** for **6 tasks** | ○ |
| 7 | You will be paid **$8.09** for **7 tasks** | ○ |
| 8 | You will be paid **$9.35** for **8 tasks** | ○ |
| 9 | You will be paid **$10.34** for **9 tasks** | ○ |
| 10 | You will be paid **$11.0** for **10 tasks** | ○ |

*Figure 4. Decision Screen*

After choosing one alternative from **each of the 20 menus,** one of the menus will be chosen at random (with equal probability). Your choice in this menu will then be implemented in Part III of the experiment. That is, you will have to finish the number of tasks corresponding to the alternative you chose in this menu. You will also receive the compensation corresponding to the option you chose.

Please make your decisions carefully. They will determine the number of tasks and the compensation you will receive in Part III.

# PART 3

Thank you for completing Part II of the experiment! We are now proceeding to the next stage of the experiment. Please read the instructions carefully. You will be able to proceed shortly after all participants finish Part II.

Please choose an alternative from each one of the next 20 menus. Once you make all your decisions, you will be presented with the selected task and the corresponding payment. Once you complete the specified tasks, you will see the final screen with your payment information. **You DO NOT NEED to wait for other participants**. Please record your **label** and quietly proceed to collect your payment.

In this part, you will complete tasks as in Part I. The system will randomly pick one of the budgets you have faced and implement the choice you have made. You will see the contract implemented on the screen, as illustrated in Figure 5.

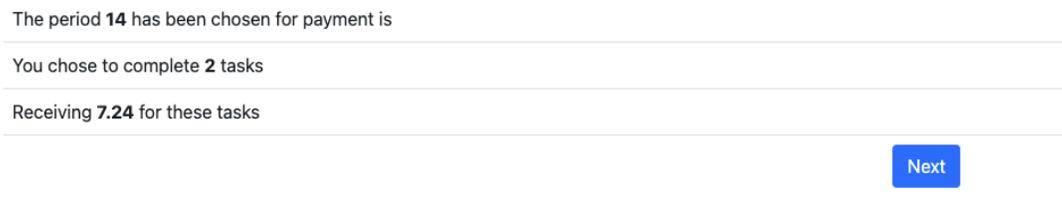

*Figure 5. Contract Choice Screen*

Recall that you will need to input the data from the scanned paper (presented on the left side of the screen) to the input form (shown on the right side of the screen), precisely as you have done in Part I. You will be paid only after completing all the tasks you have agreed to do beforehand.

Please consult the instructions of Part I for a reminder of the rules and hints for completing the tasks.

**Thank You and Good Luck!**



\end{document}